\documentclass[]{aa}  

\usepackage{xcolor}
\usepackage{comment}
\usepackage{graphicx}
\usepackage{txfonts}
\begin{document}


   \title{Vertically extended and asymmetric CN emission in the Elias 2-27 protoplanetary disk}

   \authorrunning{T. Paneque-Carreno et al.}
    \author{T. Paneque-Carre\~no \inst{1,2,3},
          A. Miotello \inst{1},
          E. F. van Dishoeck \inst{2,4},
          L. M. Pérez \inst{3,5}, \\
          S. Facchini \inst{1,6},
          A. Izquierdo \inst{1,2},
          L. Tychoniec \inst{1},
          L. Testi \inst{1}
          }

    \institute{European Southern Observatory, Karl-Schwarzschild-Str 2, 85748 Garching, Germany
    \and Leiden Observatory, Leiden University, P.O. Box 9513, NL-2300 RA Leiden, the Netherlands
    \and Departamento de Astronomía, Universidad de Chile, Camino El Observatorio 1515, Las Condes, Santiago, Chile
    \and Max-Planck-Institut für extraterrestrische Physik, Gießenbachstr. 1 , 85748 Garching bei München, Germany
    \and N\'ucleo Milenio de Formaci\'on Planetaria (NPF), Chile
    \and Dipartimento di Fisica, Università degli Studi di Milano, Via Celoria, 16, Milano, I-20133, Italy\\
    \\
              \email{tpaneque@eso.org}}
   \date{}

 
  \abstract
   {CN emission is expected to originate in the upper layers of protoplanetary disks, tracing UV-irradiated regions. This hypothesis, however, has been observationally tested only in a handful of disks. Elias 2-27 is a young star that hosts an extended, bright and inclined disk of dust and gas. The inclination and extreme flaring of the disk make Elias 2-27 an ideal target to study the vertical distribution of molecules, particularly CN.}
   {We aim to directly trace the emission of CN in the disk around Elias 2-27 and compare it to previously published CO isotopologue data of the system. Both tracers can be combined and used to constrain the physical and chemical properties of the disk. Through this analysis we can test model predictions of CN emission and compare observations of CN in other objects to the massive, highly flared, asymmetric and likely gravitationally unstable protoplanetary disk around Elias 2-27.}
   {We analyze CN $N = 3-2$ emission in two different transitions $J = 7/2 - 5/2$ and $J = 5/2 - 3/2$, for which we detect two hyperfine group transitions. The vertical location of CN emission is traced directly from the channel maps, following geometrical methods that have been previously used to analyze the CO emission of Elias 2-27. Simple analytical models are used to parametrize the vertical profile of each molecule and study the extent of each tracer. From the radial intensity profiles we compute radial profiles of column density and optical depth.}
   {We show that the vertical location of CN and CO isotopologues in Elias 2-27 is layered and consistent with predictions from thermochemical models. A north/south asymmetry in the radial extent of the CN emission is detected, which is likely due to shadowing in the north side of the disk. Combining the information from the peak brightness temperature and vertical structure radial profiles, we find that the CN emission is mostly optically thin and constrained vertically to a thin slab at $z/r \sim$0.5. A column density of 10$^{14}$\,cm$^{-2}$ is measured in the inner disk which for the north side decreases to 10$^{12}$\,cm$^{-2}$ and for the south side to 10$^{13}$\,cm$^{-2}$ in the outer regions.}
   {In Elias 2-27, CN traces a vertically elevated region above the midplane, very similar to that traced by $^{12}$CO. The inferred CN column densities, low optical depth ($\tau \leq$1) and location near the disk surface are consistent with thermo-chemical disk models, in which CN formation is initiated by the reaction of N with UV-pumped H$_2$. The observed north/south asymmetry may be caused by either ongoing infall or by a warped inner disk. This study highlights the importance of tracing the vertical location of various molecules to constrain the disk physical conditions.}


   \maketitle

%

\section{Introduction} \label{sec:intro}

Protoplanetary disks are the formation sites of planetary systems. Studying the dust and gas components of these disks is crucial for understanding the origin and composition of planets \citep{2020ARA&A..58..483A}. Analyzing the radial and vertical distribution of molecular tracers within a protoplanetary disk offers spatially located information on the temperature gradient, abundances and physical conditions of a system \citep[e.g.][]{MAPS_law_radial, MAPS_4_height_law, MAPS_vivi}. To conduct these studies, high spectral and spatial resolution observations are required, a task that has been possible in recent years with the Atacama Large Millimeter/submillimeter Array (ALMA). High resolution studies so far have been biased towards either a few of the brightest and largest sources in the dust continuum \citep{DSHARP_Andrews, MAPS_1_oberg}, and/or towards studying the most abundant and brightest molecular tracers, mostly common CO isotopologues \citep{2018ApJ...859...21A, 2017ApJ...844...99L, 2016ApJ...827..142B, 2017A&A...599A.113M, 2014A&A...564A..95P, 2015A&A...579A.106V, 2016A&A...585A..58V}. Less abundant molecules have also been studied in some sources, at lower spatial and/or spectral resolution \citep{2018ApJ...863..106K, 2018ApJ...857...69B, 2019A&A...631A..69M, 2019ApJ...881..127A, 2019ApJ...876...25B,  2019ApJ...876...72L, 2021A&A...645A.145G, MAPS_ilee, MAPS_vivi}. These studies have shown the potential of other tracers to give us information on not only chemical abundance and spatial location, but also on temperature profiles, density, ionization, atomic ratios and kinematics \citep[see ][for a complete overview and review]{MAPS_1_oberg, 2021PhR...893....1O}.

Studying various molecular tracers is key towards completing our understanding of the disk structure, which can be vertically divided in three distinct zones; 1) a hot atmosphere mainly populated by atomic gas, heavily irradiated by energetic UV photons and X-rays from the star, as well as cosmic rays from the environment, 2) a warm molecular layer, where most of the chemical interactions occur and molecules may reside in gas phase, 3) a cold midplane, where larger grains of dust are found and some simple molecules may be present in gas phase, however, most of the molecules will be found in ice form, coating the dust grains \citep{2002A&A...386..622A, 2007prpl.conf..751B, 2013ChRv..113.9016H, 2014prpl.conf..317D}. To gain information on the whole disk structure, we must characterize the emission from molecules residing in the various layers \citep[e.g.][]{2001A&A...377..566V, MAPS_4_height_law}.

Cyanide (CN) is a molecule that is abundant enough to be bright and observable in relatively short integration times \citep{2008A&A...492..469K, 2019A&A...623A.150V}. Theoretically its main formation pathway is the reaction of N with vibrationally excited H$_{2}$, through UV pumping of H$_{2}$, which occurs in the uppermost regions of the disk atmosphere \citep{2018A&A...609A..93C, 2018A&A...615A..75V}. This enables CN to be a unique tracer of the upper disk structure and UV radiation from photons of wavelengths $<105$\,nm \citep{MAPS_CN_Bergner, 2017A&A...602A.105H}. CN also has multiple hyperfine structure transitions, which come from the splitting of energy levels by nuclear magnetic interactions. Studying the hyperfine structure components allows us to directly determine various physical properties, such as the density and optical depth of the emission \citep[e.g.,][]{1999ApJ...517..209G, 2018ApJ...859..131L, Facchini_2021}. The study of multiple CN hyperfine transitions has been possible in TW Hya \citep[][]{2014ApJ...793...55K, 2016A&A...592A..49T, 2020ApJ...899..157T}, V4046\,Sgr \citep{2014ApJ...793...55K} and more recently, in HD\,163296 \citep{MAPS_CN_Bergner}.

Individual analyses of resolved sources have shown ring-like CN emission in disks \citep{2019A&A...623A.150V, 2020ApJ...899..157T, MAPS_CN_Bergner}. Theoretically, this has been predicted by \cite{2018A&A...609A..93C} as a consequence of CN formation mediated by the balance of excited H$_{2}$, via FUV pumping, and the collisional de-excitation of H$_{2}$ . Most recently, using models of UV-radiation, \cite{MAPS_CN_Bergner} show that the radial extent and morphology of the CN emission in various systems is in agreement with the variations in the UV flux from photons with wavelengths of 91-102\,nm. Observational studies are consistent with theoretical predictions that CN emission traces zones of UV radiation, regions at about $z/r \sim 0.2 - 0.4$ \citep{2020ApJ...899..157T, 2021A&A...646A..59R, MAPS_CN_Bergner}. However, the height of the emission in observations has been mostly determined indirectly through models \citep{MAPS_CN_Bergner} or temperature estimates \citep{2020ApJ...899..157T}. To the best of our knowledge, a direct measurement of the CN emitting layer in a protoplanetary disk of a Class II stellar source has only been done in the edge-on system known as the Flying Saucer \citep{2021A&A...646A..59R}.

In this study we present new CN observations of the disk surrounding Elias 2-27, a Class II young stellar object \citep{2009ApJ...700.1502A, DSHARP_Andrews} which has been proposed to host a massive \citep[$\sim$17\% of the stellar mass,][]{veronesi_mass} protoplanetary disk currently undergoing gravitational instability (GI) \citep{laura_elias, DSHARP_Huang_Spirals, paneque_elias1}. Elias 2-27 is located in the star-forming region of $\rho$ Oph, at 115.8\,pc distance \citep{2018A&A...616A...1G}. Dust continuum observations at multiple wavelengths show a symmetric, double spiral morphology \citep{laura_elias, DSHARP_Huang_Spirals, paneque_elias1}, with signatures of dust trapping along the spirals \citep{paneque_elias1}. Spectral line studies have been challenging, even with bright tracers like isotopologues of CO $J = 2-1$, due to the heavy cloud absorption that affects the system \citep{laura_elias, DSHARP_Andrews, 2020ApJ...890L...9P}. For this reason, previous attempts to observe CN $N = 2 - 1$ were unsuccessful \citep{2015A&A...578A..31R}. Recently, \cite{paneque_elias1} presented C$^{18}$O and $^{13}$CO $J = 3-2$ data that are much less affected by absorption, showing that higher energy transitions may be used to characterize the gas structure and kinematics of the source. 

\begin{figure*}
   \centering
   \includegraphics[width=\hsize]{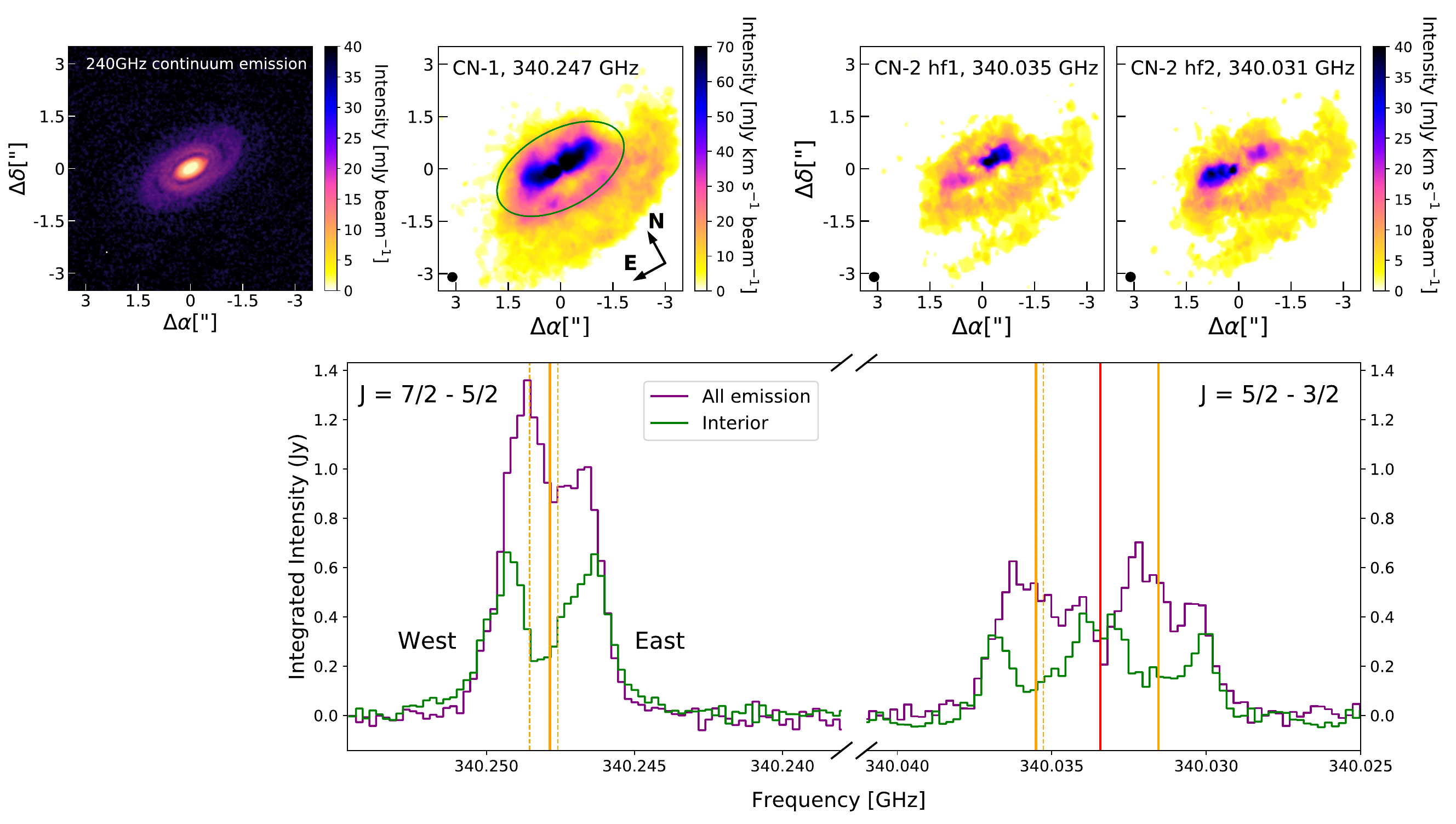}
      \caption{Top row: left corner displays the high-resolution 240\,GHz dust continuum emission. The subsequent panels towards the right show the integrated intensity (zeroth moment) maps for CN $J = 7/2 - 5/2$ (CN-1) and CN $J = 5/2 - 3/2$ hyperfine group transitions at $\sim$340.035\,GHz (CN-2 hf1) and $\sim$340.031\,GHz (CN-2 hf2) in the two rightmost panels. The frequency of each group is shown in the top of each panel and the ellipses in the bottom left corners show the beam size. In the left panel the green ellipse encircles the region within 230\,au from the center (interior region), assuming a flat geometry. Bottom row: Disk integrated spectrum for the whole emitting region in purple and only for the interior region in green. The orange vertical lines mark the hyperfine transition frequencies that are blended in each peak. For those with multiple transitions, the solid vertical line indicates the one with higher relative intensity (values in Table \ref{table_molec}). The red vertical line marks the separation between the CN-2 hyperfine groups
              }
         \label{obs_CN_spectra}
\end{figure*}

\begin{table*}[h]
\def\arraystretch{1.5}
\setlength{\tabcolsep}{7pt}
\caption{Molecular line data, values from CDMS catalogue \citep{2001A&A...370L..49M}}
\label{table_molec}      
\centering
\begin{tabular}{c c c c c c c c c}       
\hline\hline                
Species & $J$ Transition & $F$ Transition& Frequency & $E_u$ &  $A_{ul}$ & $g_u$ & R.I.\tablefootmark{*} & Int. Flux \tablefootmark{**}\\   
 &  & & (GHz)& (K) & (s$^{-1}$ )& & & (Jy\,km\,s$^{-1}$)\\   
\hline                       
    
   $^{12}$CO & $J=2-1$ & & 230.5380000 & 16.59 & 6.91$\times$10$^{-7}$ & 5 & &29.61 $\pm$ 1.28\\
   $^{13}$CO & $J=3-2$ &  &330.5879653  & 31.73 &2.19$\times$10$^{-6}$ & 14 & & 22.26 $\pm$ 0.09\\ 
   
   C$^{18}$O  & $J=3-2$ &  &329.3305525  & 31.61 &2.17$\times$10$^{-6}$ & 7 &  & 4.45 $\pm$ 0.06\\
\hline  
   CN-1 & $J=7/2-5/2$  &  $F=7/2-5/2$ &340.247590\tablefootmark{a}  & 32.67 &3.797$\times$10$^{-4}$   & 8 & 0.307 & 3.55 $\pm$ 0.07\\ 
    &   & $F=9/2-7/2$ & 340.247861\tablefootmark{a} & 32.67 &4.13$\times$10$^{-4}$  & 10 & 0.417 &\\ 
     &   & $F=5/2-3/2$ & 340.248544 &32.67 &3.67$\times$10$^{-4}$  & 6 & 0.222 &\\ 
\hline
   CN-2  hf1 & $J=5/2-3/2$ & $F=3/2-1/2$&  340.035269\tablefootmark{b} & 32.63 &2.89$\times$10$^{-4}$  & 4 & 0.167 & 1.49 $\pm$ 0.08\\
   &  & $F=5/2-3/2$&  340.035507\tablefootmark{b}  &  32.63 &3.23$\times$10$^{-4}$  & 6 & 0.281 &\\
\hline
   CN-2 hf2 & $J=5/2-3/2$& $F=7/2-5/2$ & 340.0315494  &  32.63 &3.85$\times$10$^{-4}$ & 8 & 0.445 & 1.62 $\pm$ 0.08\\
\hline      
\end{tabular}
\tablefoot{
\tablefoottext{*}{Relative Intensity values for hyperfine CN transitions obtained from \citet{2017A&A...603L...6H} and references therein.}
\tablefoottext{**}{Integrated flux values and their uncertainties for the CN lines calculated from the emission spectra in Figure \ref{obs_CN_spectra} and in the same way for CO isotopologues from Figure \ref{obs_CO_spectra}}
\tablefoottext{a}{Value in CDMS catalog is 340.2477700 for both, new value taken from \citet{2020ApJ...899..157T} }
\tablefoottext{b}{Value in CDMS catalog is 340.0354080 for both, new value taken from \citet{2020ApJ...899..157T} }
}
\end{table*}

Thanks to the inclination of the system (56.2$^{\circ}$, \citealt{DSHARP_Huang_Annular}) Elias 2-27 is a prime target to study both the radial distribution and the vertical location of any molecular tracer. A study of the vertical structure of Elias 2-27 was done using C$^{18}$O and $^{13}$CO data by \cite{paneque_elias1}, directly tracing the emission from the channel maps and applying geometrical methods to obtain the emitting surface \citep[following the method outlined in ][]{pinte_2018_method}. In the CO isotopologue data an asymmetry in the vertical structure between the east and west sides of the disk was found and the origin of this variation was proposed to be ongoing infall \citep{paneque_elias1}. We expect CO isotopologues to trace the intermediate layer of the disk \citep{2007prpl.conf..751B, 2014prpl.conf..317D} and CN to contain information on the upper layers of the disk \citep{2018A&A...609A..93C}. 

Using the disk structure as traced with CN we can compare the distribution of CN against the asymmetries found in CO isotopologues. The constraints provided on the location of the CN emission in Elias 2-27 can be compared with theoretical models of CN \citep{2018A&A...609A..93C} and previous observations of other systems \citep{2020ApJ...899..157T, 2021A&A...646A..59R, MAPS_CN_Bergner}. Finally, combining the information on the emission and location of the available tracers, we can study the overall temperature, column density and distribution of molecules in the source. 

As Elias 2-27 is the first system to show strong evidence, in multiple tracers, of being gravitationally unstable \citep{laura_elias, paneque_elias1} it is also an excellent test-case for predictions of GI chemistry \citep{2011MNRAS.417.2950I, 2015MNRAS.453.1147E}. In particular, CN has been predicted to be a good tracer of GI due to the temperatures along GI spirals. The temperature in the spirals depends on how massive the disk is, but can reach values of up to $\sim$400\,K \citep[inner $\sim$20\,au,][]{2015MNRAS.453.1147E}, at these high temperatures CN would be destroyed due to its reaction with NH$_3$ \citep{2015MNRAS.453.1147E} or H$_2$ \citep{2018A&A...615A..75V}. However, in the outer parts of the disk, the spirals may have more moderate temperatures and reach values that would lead to CN being desorbed from the dust grains in the midplane. The desorption temperature of CN is around 120\,K \citep{2015MNRAS.453.1147E} therefore, if this temperature is reached and CN is emitted from the midplane, it may trace the spiral arms as seen in the dust continuum.

The remainder of the paper is distributed as follows, Section 2 describes the imaging procedure and main observed features. Section 3 details the analysis of the vertical location of the CN emission, comparing it to the available CO tracers in Elias 2-27. Section 4 presents results obtained from CN emission and we study the asymmetries in the emission extent of the disk. Section 5 describes the derived temperature structure, abundance and optical depth of the disk emission. Section 6 discusses the implications of our findings and section 7 lists our main conclusions.

\section{Observations} \label{sec:obs}
\subsection{Self calibration and Imaging}

We present the analysis of CN $v = 0$, $N = 3 - 2$ in $J = 7/2 - 5/2$ ($E_{u} = 32.67$\,K) and $J = 5/2 - 3/2$ ($E_{u} = 32.63$\,K) emission of the disk around Elias 2-27. The data are part of ALMA programs \#2016.1.00606.S and \#2017.1.00069.S, and were obtained simultaneously with C$^{18}$O and $^{13}$CO ($J = 3 - 2$) spectral line observations presented in \cite{paneque_elias1} and also used in this work. The calibration, imaging and initial analysis of C$^{18}$O and $^{13}$CO is described in \cite{paneque_elias1}. Additionally, we use archival $^{12}$CO $J = 2-1$ data obtained as part of the DSHARP program \citep{DSHARP_Andrews}. Calibration and imaging details for $^{12}$CO can be found in the original DSHARP papers \citep{DSHARP_Andrews, DSHARP_Huang_Spirals}.

The spectral windows for the CN $v = 0$, $N = 3 - 2$ observations are centered at 340.248\,GHz ($J = 7/2 - 5/2 $) and 340.014\,GHz ($J = 5/2 - 3/2 $), respectively, with a spectral resolution of 0.2\,MHz. For the $J = 5/2 - 3/2$ data we measure two peaks of line emission in the border channels, at 340.035\,GHz and 340.031\,GHz. This corresponds to the blended emission from hyperfine transitions $F = 3/2-1/2$ and $F = 5/2-3/2$ (340.035\,GHz) and emission from hyperfine transition $F = 7/2-5/2$ (340.031\,GHz). The CN $J = 5/2 - 3/2$ fine structure group has two other hyperfine components that we do not detect, $F = 5/2-5/2$ (340.008\,GHz) and $F = 3/2-3/2$ (340.019\,GHz) which are the transitions with lowest relative intensities within the group \citep[0.054 each, where 1 is the total intensity of the group,][and references therein]{2017A&A...603L...6H}. In the CN $J = 7/2 - 5/2 $ group, the emission we measure is the blended contribution from hyperfine components $F=7/2-5/2, F=9/2-7/2$ (both 340.247\,GHz) and $F=5/2-3/2$ (340.248\,GHz). In this group we do not detect emission from the lowest relative intensity components $F=5/2-5/2$ (340.261\,GHz) and $F=7/2-7/2$ (340.265\,GHz), \citep[relative intensity of 0.027 each, where 1 is the total intensity of the group,][and references therein]{2017A&A...603L...6H}. As the CN data were obtained simultaneously with Band 7 continuum (0.89\,mm) observations, we apply the self calibration solutions of that continuum dataset to the CN measurements \citep[for details see][]{paneque_elias1}. 

Previous CO data show that Elias 2-27 presents extended large-scale emission \citep{paneque_elias1}. We check for any large-scale structure in the CN transitions by imaging the whole field of view available for CN using a robust parameter of 2.0. We do not detect any large scale emission up to a sensitivity of 2.2\,mJy\,beam$^{-1}$. Even with the non-detection of large-scale structure, we perform the final CN imaging using \texttt{uvrange} to ensure we are imaging emission only from disk scales. Additionally, \texttt{uvtaper} is applied to obtain a roughly round beam, which will facilitate the analysis and comparison between transitions.

CN $J = 7/2 - 5/2$ emission is imaged with a robust parameter of 1.3 and we use \texttt{uvrange} to only consider baselines larger than 34\,m, excluding any possible emission from scales larger than 6.5$\arcsec$ (750\,au, at 115\,pc). We apply \texttt{uvtaper} of  0.175$\arcsec \times$ 0$\arcsec$ and obtain a final beam of 0.29$\arcsec \times$0.29$\arcsec$ (34\,au, at 115\,pc). CN $J = 5/2 - 3/2$ is imaged with robust parameter 1.8, applying \texttt{uvrange} to use only baselines over 35\,m (excluding emission larger than 6$\arcsec$) and with \texttt{uvtaper} of 0.18$\arcsec \times$0.01$\arcsec$ to achieve a final beam of 0.3$\arcsec \times$ 0.3$\arcsec$ (35\,au, at 115\,pc).

\begin{figure*}[h!]
   \centering
   \includegraphics[width=\hsize]{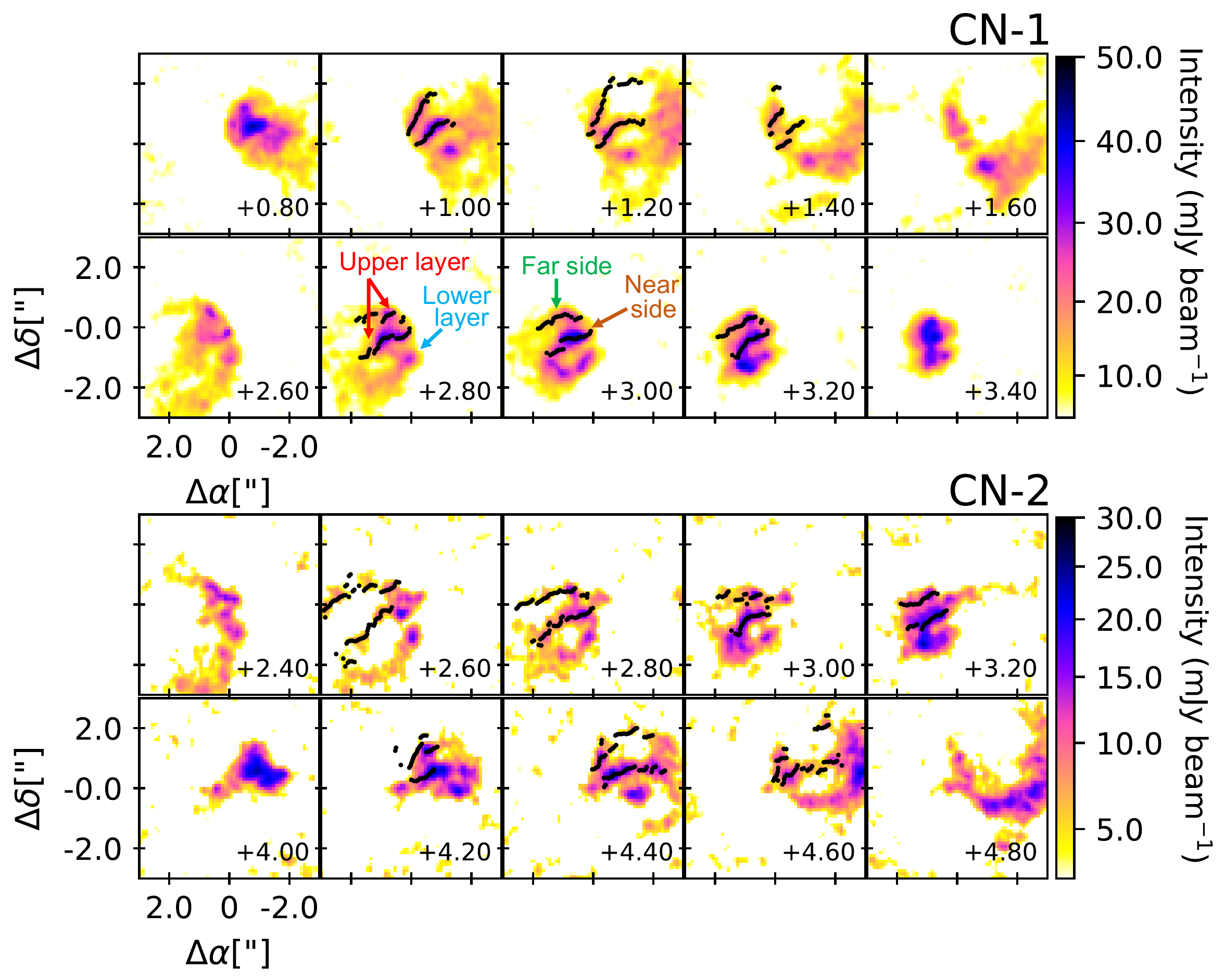}
      \caption{Panels showing a selection of channel maps for CN-1 (top) and CN-2 (bottom) emission. The numbers on the top right corners in each panel show the channel velocity (systemic velocity is 1.95\,km\,s$^{-1}$) and the ellipses on the bottom left corners the beam size. For the channels used to derive the vertical profile, black dots trace the peak intensity locations along the upper layer of emission. Upper and lower layers of emission, as well as far and near sides of the emission are identified for reference in the bottom panels of CN-1 emission. A complete panel of the channel map emission is presented in Appendix A for each transition.
              }
         \label{channels_CN_max}
\end{figure*}

\subsection{Emission morphology and line ratios}

For easier identification, hereafter we refer to the blended CN $J = 7/2 - 5/2$ emission as CN-1 and to the $J = 5/2 - 3/2$ group (two emission peaks) as CN-2, labeling with hf1 and hf2 for the blended emission at 340.035\,GHz and 340.031\,GHz, respectively. The integrated intensity maps of each CN transition are shown in the top row of Figure \ref{obs_CN_spectra}, together with a 240\,GHz high resolution continuum image of Elias 2-27 \citep{DSHARP_Andrews}. The bottom row shows the integrated spectra of the observations. This figure shows that the CN emission has an asymmetrical distribution between north and south sides of the disk in all transitions. The emission is centrally peaked and elongated along the major axis, however, towards the north there is not much emission beyond the bright inner region. In the south side of the disk, the emission extends up to $\sim$2\,$\arcsec$ in the sky plane from the center and shows arc-like features. The CN-2 integrated emission (mom0) maps have been computed by visual identification of the two hyperfine lines in the integrated spectrum and setting a separation between both of them at 340.033\,GHz (see red line in Fig. \ref{obs_CN_spectra}, bottom panel). Relative to their respective central frequencies, CN-2 hf1 lacks redshifted emission (spatially located in east side) and CN-2 hf2 misses blueshifted emission (spatially located in west side). The differences observed in the central emission between east and west sides of CN-2 hf1 and hf2 (top right panels of Figure \ref{obs_CN_spectra}) are likely artificial and due to the channel selection when computing the zeroth moment map.

Two spectra are displayed for each transition group (bottom panel in Figure \ref{obs_CN_spectra}). The purple spectrum considers the complete emitting region of the disk, the green spectrum is obtained from the region within 230\,au of the star, considering a flat geometry. In all groups, the purple spectrum shows a higher intensity towards redshifted frequencies with respect to the central frequency of the strongest transition (marked by the solid orange vertical line). This peak of emission corresponds spatially to the west side of the disk and the asymmetry is not seen in the spectrum from the interior region. We can visually assess that beyond the interior region there is further emission in the south of the disk that is more azimuthally extended towards the west side. Therefore, the flux difference in the spectrum is tracing an emission asymmetry between the east and west sides of the disk at large radii. From CO observations it is known that Elias 2-27 is largely affected by cloud absorption on the east side of the disk \citep{laura_elias, DSHARP_Huang_Spirals}, however, we do not see a flux difference in the CN emission when considering only the interior region. We hence consider it is unlikely that the flux asymmetry of CN at large scales ($>$230\,au) is caused by cloud absorption. The visually identified emission extension difference between north and south will be studied in detail in Section 4.

Table \ref{table_molec} shows the molecular line data from the literature and measured integrated fluxes for the detected CN transitions and CO lines that will be used in the remainder of this work. In Appendix B we explain the procedure to determine the integrated flux and associated errors for each tracer and the integrated emission maps and spectra are presented for each CO line (Figure \ref{obs_CO_spectra}). We compare the integrated flux ratios of the CN emission to the expected ratio values, obtained from the relative intensities \citep[R.I., see table A.1 in ][]{2017A&A...603L...6H} and the Einstein emission coefficient ($A_{ul}$) of each hyperfine component. The R.I. value is normalized so that within each fine structure group the total sum of all individual values will equal unity \citep{2017A&A...603L...6H}. Our table shows the R.I. values only for the transitions that are detected, therefore the values within each group do not sum unity. We have separated the CN-2 emission in hf1 and hf2, however they are both from the same fine structure group $J = 5/2 - 3/2$ and can be easily compared. The emission labeled as hf1 has two blended components, their R.I. sum to 0.448. CN-2 hf2 has only one hyperfine transition, that accounts for 0.445 of the group intensity. Therefore, the expected intensity ratio between both components is hf1/hf2 $=$ 0.448/0.445 $=$ 1.01. From the integrated emission values we calculate that hf1/hf2 $=$ 1.49/1.62 $=$ 0.92$\pm$0.07 (error is obtained considering the propagation of uncertainties listed in Table \ref{table_molec}). The ratio calculated from observational constraints is very close to the theoretical prediction and suggestive of CN-2 being optically thin emission. 

To compare CN-1 with CN-2 we take into account the ratio between Einstein coefficients, as the emission is from different fine structure groups \citep{2017A&A...603L...6H}. The total relative intensity for the blended CN-1 emission is 0.946. To calculate the expected intensity ratio with CN-2 hf2 we calculate (0.946/0.445)(4.13$\times$10$^{-4}$/3.85$\times$10$^{-4}$), where the second term is the Einstein coefficient ratio. For multiple blended transitions, we use the coefficient value of the hyperfine component with the highest relative intensity in the group. The expected ratio of CN-1 with CN-2 hf2 is then 2.28 and for CN-1 with the full CN-2 emission (both peaks) the expected ratio is 1.14. From the integrated intensity, the observational ratios are 2.19$\pm$0.12 and 1.14$\pm$0.05. As in the comparison between CN-2 peaks, the theoretical and observational values are very similar, indicating that overall CN emission is likely to be optically thin. It is important to note that when the flux ratios are close to unity they do not provide a strong discriminant between optically thick and thin regimes, even if the theoretical prediction matches the observed values, as is the case here. In section 5 we present an independent way to determine the optical depth and physical properties of the CN emission.

\begin{figure*}[h!]
   \centering
   \includegraphics[scale=0.65]{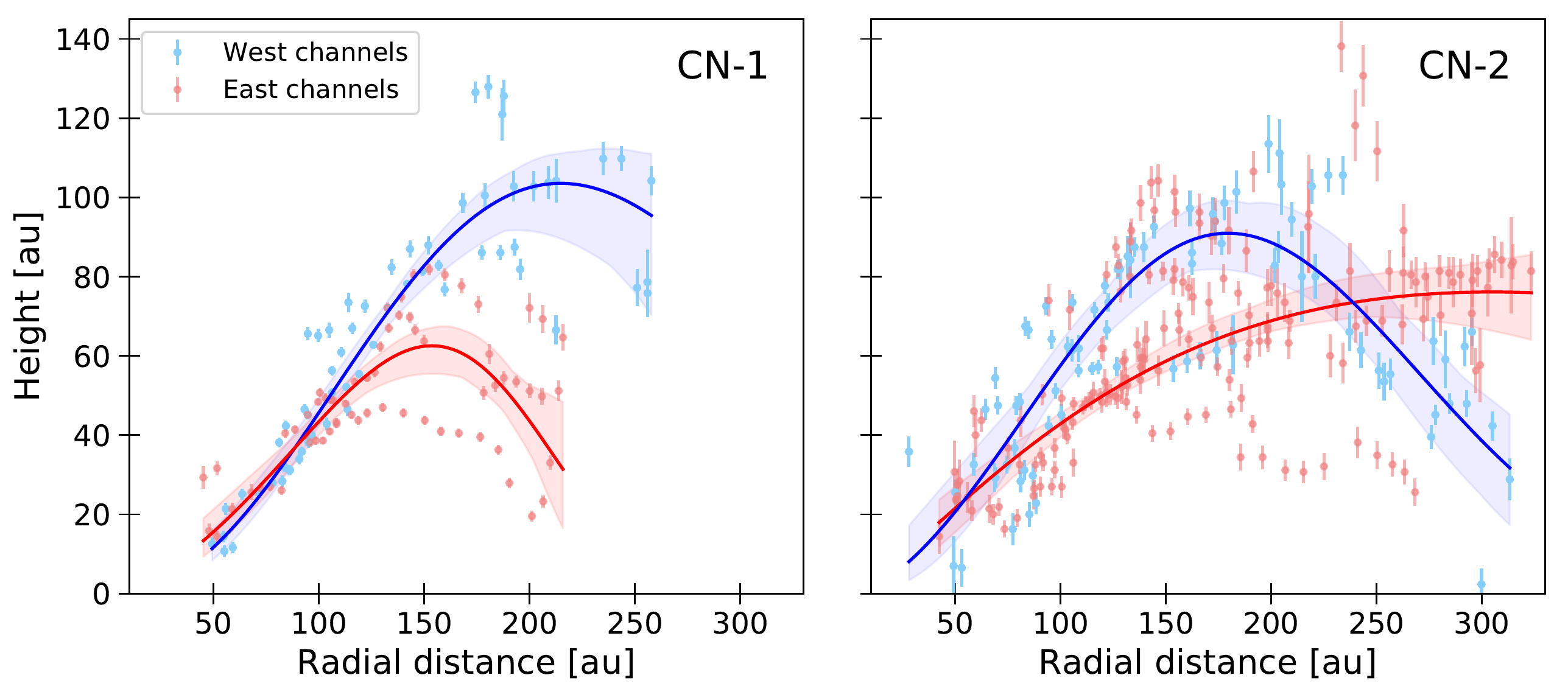}
      \caption{Vertical location of the emission from CN-1 (left) and CN-2 (right) transitions, using data obtained from the channel map analysis of the emission. The data set is separated between East (red dots) and West (blue dots) sides of the disk, as defined by the semi-minor axis. The solid blue and red curves trace the best-fit exponentially tapered model of each side. The colored regions show the uncertainty of the model, obtained from the spread of the posterior values from our MCMC analysis.}
         \label{height_CN_best}
\end{figure*}

\begin{figure*}[h!]
   \centering
   \includegraphics[scale=0.65]{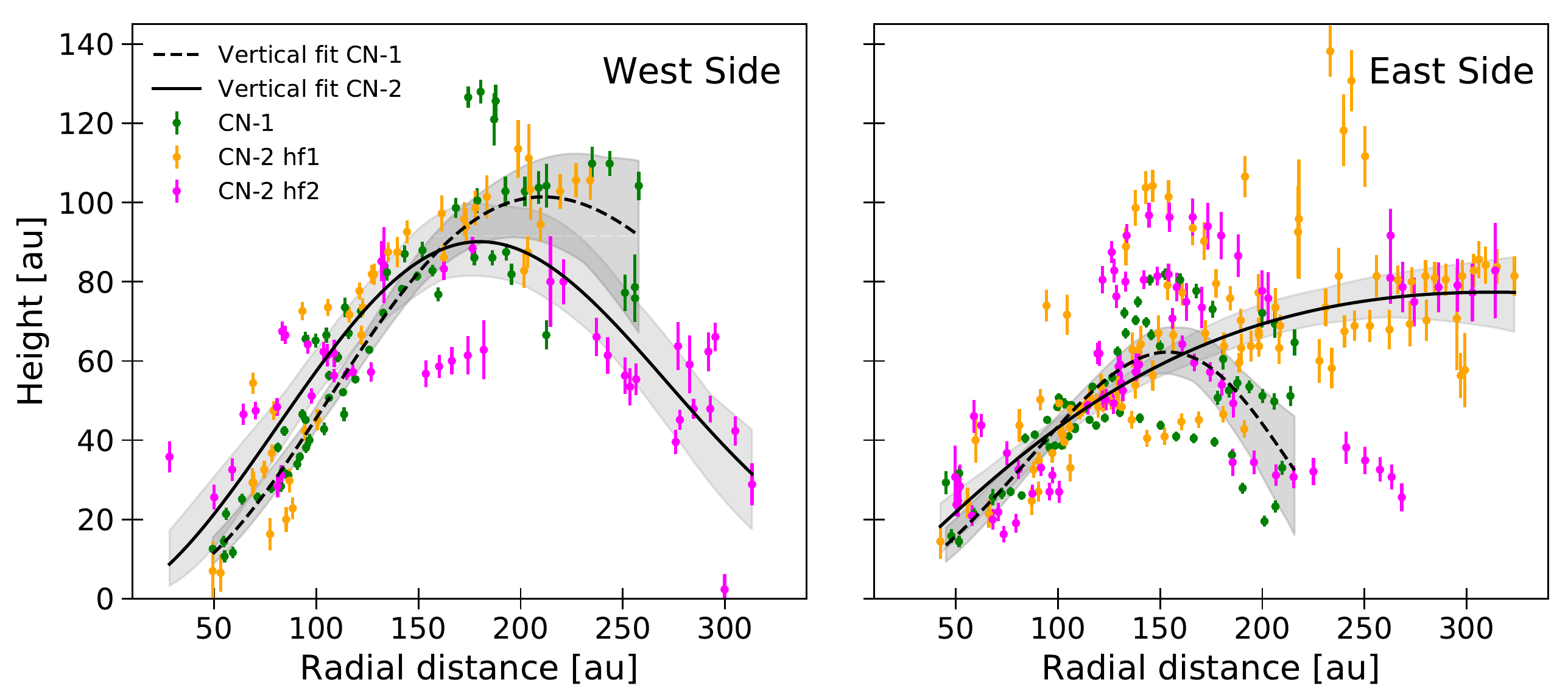}
      \caption{Same data as shown in Figure \ref{height_CN_best}, but the panels are now separated by west and east sides of the disk. Colored data shows the identified CN hyperfine transitions. Solid and dashed black curves show the best fit vertical models for CN-1 and CN-2, respectively. The light grey region identifies the uncertainties in the best-fit vertical profile, obtained from the spread of the posterior values of the MCMC model.}
         \label{height_CN_sides}
\end{figure*}

\section{Vertical structure analysis}

\subsection{Height of CN emission}

A selection of the channel maps showing the emission from each CN transition is displayed in Figure \ref{channels_CN_max}. We note that for CN-2 the top row shows emission from hf1 and the bottom row from hf2 transitions. For the complete panel with all available channels see Figures \ref{panel_CN1} and \ref{panel_CN2} for CN-1 and CN-2, respectively. The systemic velocity of the system is determined to be $V_{LSR} \sim$1.95\,km\,s$^{-1}$\citep{paneque_elias1}. From the channel maps we identify that the emission does not come from the midplane, but rather originates from a constrained and flared vertical layer. This can be deduced from the separation between the emission from lower and upper layers, which are visually identified (see Figure \ref{channels_CN_max}) and indicate the vertical location of the gas emission with respect to the disk midplane, as observed from our line-of-sight. These layers can be visually identified from the channel maps aided by the prior information on the CO distribution \citep{paneque_elias1} and the geometry of the system \citep{DSHARP_Huang_Spirals}. To extract the radial vertical profile we apply the method developed by \cite{pinte_2018_method}, which directly traces the emitting surface from the channel maps. This is done by following the maxima of emission along the upper layer in channels where it is possible to identify far and near sides of the upper layer \citep[for a detailed sketch see Figures 2 and 3 of][]{pinte_2018_method}.
This method has been previously used for the extraction of the C$^{18}$O and $^{13}$CO vertical emitting layer in Elias 2-27 \citep{paneque_elias1} and in various works, for other sources and/or CO isotopologues \citep[e.g.][]{MAPS_law_radial, 2021ApJ...913..138R, Leemker_2022}.

\begin{table*}[h!]
\def\arraystretch{1.5}
\setlength{\tabcolsep}{5pt}
\caption{Best-fit model parameters of exponentially tapered vertical profile from channel analysis. Data presented in Figure \ref{height_CN_best}.}
\label{table_param_height}      
\centering
\begin{tabular}{c c c c c c }

\hline\hline                
Transition & side & $z_0$ [au] & $\phi$  & $r_{\mathrm{taper}}$ [au] & $\psi$  \\    
\hline                       
          
   CN-1 & W & 59.5$^{+16.78}_{-9.07}$ & 2.25$^{+0.24}_{-0.27}$ & 196.53$^{+35.86}_{-38.39}$ & 1.96$^{+0.69}_{-0.47}$\\ 
   
   CN-1 & E & 46.0$^{+4.54}_{-2.15}$ & 1.54$^{+0.21}_{-0.14}$ & 195.2$^{+6.16}_{-11.18}$ & 4.27$^{+1.02}_{-1.08}$  \\
   
   CN-2 & W & 66.13$^{+12.1}_{-5.79}$ & 1.59$^{+0.43}_{-0.29}$ & 216.37$^{+28.37}_{-38.53}$ & 2.54$^{+0.65}_{-0.53}$ \\ 
   
   CN-2 & E & 67.77$^{+13.61}_{-14.5}$ & 1.3$^{+0.23}_{-0.2}$ & 229.03$^{+124.85}_{-66.01}$ & 0.96$^{+0.29}_{-0.14}$ \\
\hline 

\end{tabular}

\end{table*}

\begin{figure*}[h!]
   \centering
   \includegraphics[scale=0.55]{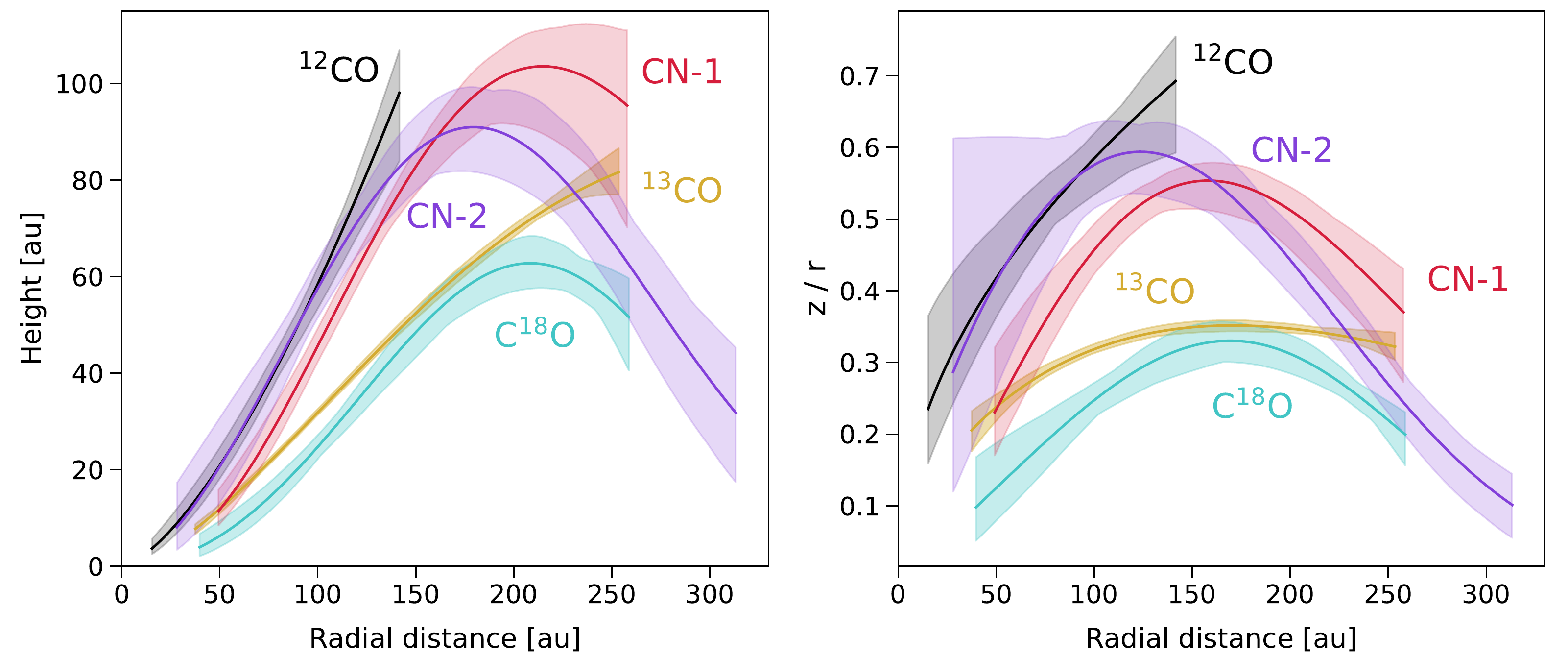}
      \caption{ Height (left panel) and $z/r$ (right panel) profiles for the available CO isotopologues and CN transitions in Elias 2-27. The CN emission is separated by hyperfine group. Solid lines trace the best-fit exponentially tapered vertical profile and the shadowed regions show the uncertainty on the best-fit model, derived from the spread of the posteriors in each tracer. Due to the absorption and asymmetries that affect Elias 2-27 the data corresponds to measurements only from the west side of the disk (for all tracers). 
              }
         \label{comp_CN_CO}
\end{figure*} 

The channels used for this analysis are those between (and including) 1.0-1.4\,km\,s$^{-1}$ and 2.8-3.2\,km\,s$^{-1}$ in CN-1, 1.0-1.2\,km\,s$^{-1}$ and 2.6-3.2\,km\,s$^{-1}$ for CN-2 hf1 and 4.2-4.6\,km\,s$^{-1}$ and 6.0-6.4\,km\,s$^{-1}$ for CN-2 hf2. The selection of channels was done through visual inspection, choosing those where it was possible to identify the upper layer on the near and far sides. Each selected channel is rotated to correct for the disk position angle and to work in cartesian coordinates for the extraction of the emission maxima \citep[for details refer to Figure 2 of][]{pinte_2018_method}. From the rotated channel emission, we mask the near and far sides of the upper layer. The pixel of peak intensity is automatically identified from the pixels within the mask, sampling every quarter of the beam along the disk major axis. Figure \ref{channels_CN_max} shows the retrieved peak intensity values (black dots) of the upper layer in some of the selected channels. The height of the emission at a given radial distance is retrieved by applying geometrical relations to the cartesian location of the emission maxima \citep[for details refer to][]{pinte_2018_method}.

It has been shown for the CO isotopologues that the emission from the east side of the disk appears to be coming from a layer closer to the midplane than the west side of the disk \citep{paneque_elias1}. Considering the latter, we divide the data analysis for CN between east and west sides of the disk, as defined by the projected minor axis. Figure \ref{height_CN_best} shows the extracted vertical profiles of each CN transition. The data are fitted using an exponentially tapered power law defined as,

\begin{equation}
z(r) = z_0\times \left(\frac{r}{100\,\mathrm{au}}\right)^{\phi} \times \exp\left[\left(\frac{-r}{r_{\mathrm{taper}}}\right)^{\psi}\right]
\end{equation}

where $r$ is the radial distance from the star along the midplane. The best fit parameters of the final model are obtained from the median value of the posteriors using a Markov chain Monte Carlo (MCMC) simulation \citep{2010CAMCS...5...65G} as implemented by \texttt{emcee} \citep{2013PASP..125..306F}. The best-fit values of each parameter and its uncertainties (16th and 84th percentile uncertainties derived from the posteriors) are shown in Table \ref{table_param_height}.

Figure \ref{height_CN_sides} shows the extracted data points separated by disk sides and hyperfine structure. It is apparent that the data from the west side of the disk are coherent between the various transitions and the best-fit vertical profiles of CN-1 and CN-2 trace the same region, within their uncertainties. The vertical structure from the east side of the disk, shows a large scatter in the extracted data, which makes the results vary largely between CN-1 and CN-2 emission. The scatter may be caused by environment conditions, as the signatures of absorption and ongoing infall from CO emission are all in the east side of the disk \citep{paneque_elias1}. Considering the well defined surface of the west side in all CN transitions and the lack of a strong east/west asymmetry, for the remainder of this work the best-fit vertical profile from the west will be used as representative of the CN distribution in the whole disk.

\subsection{CN vs CO emission}

The vertical distribution of the emission from different molecular layers in Elias 2-27 allows us to characterize the various chemical processes occurring in the disk and to estimate the 2D temperature structure of the system \citep[e.g.][]{pinte_2018_method, MAPS_4_height_law}. We compare the CN tracers with previous CO isotopologue observations \citep{paneque_elias1} and additionally present the results from extracting the vertical structure of the high-resolution $^{12}$CO $J = 2-1$ DSHARP data \citep{DSHARP_Andrews}. We do this comparison only in the west side, due to the cloud contamination that heavily affects the east side in all CO tracers, particularly $^{12}$CO $J = 2-1$ \citep{laura_elias, DSHARP_Huang_Spirals}. 

\begin{figure}[h!]
   \centering
   \includegraphics[scale=0.73]{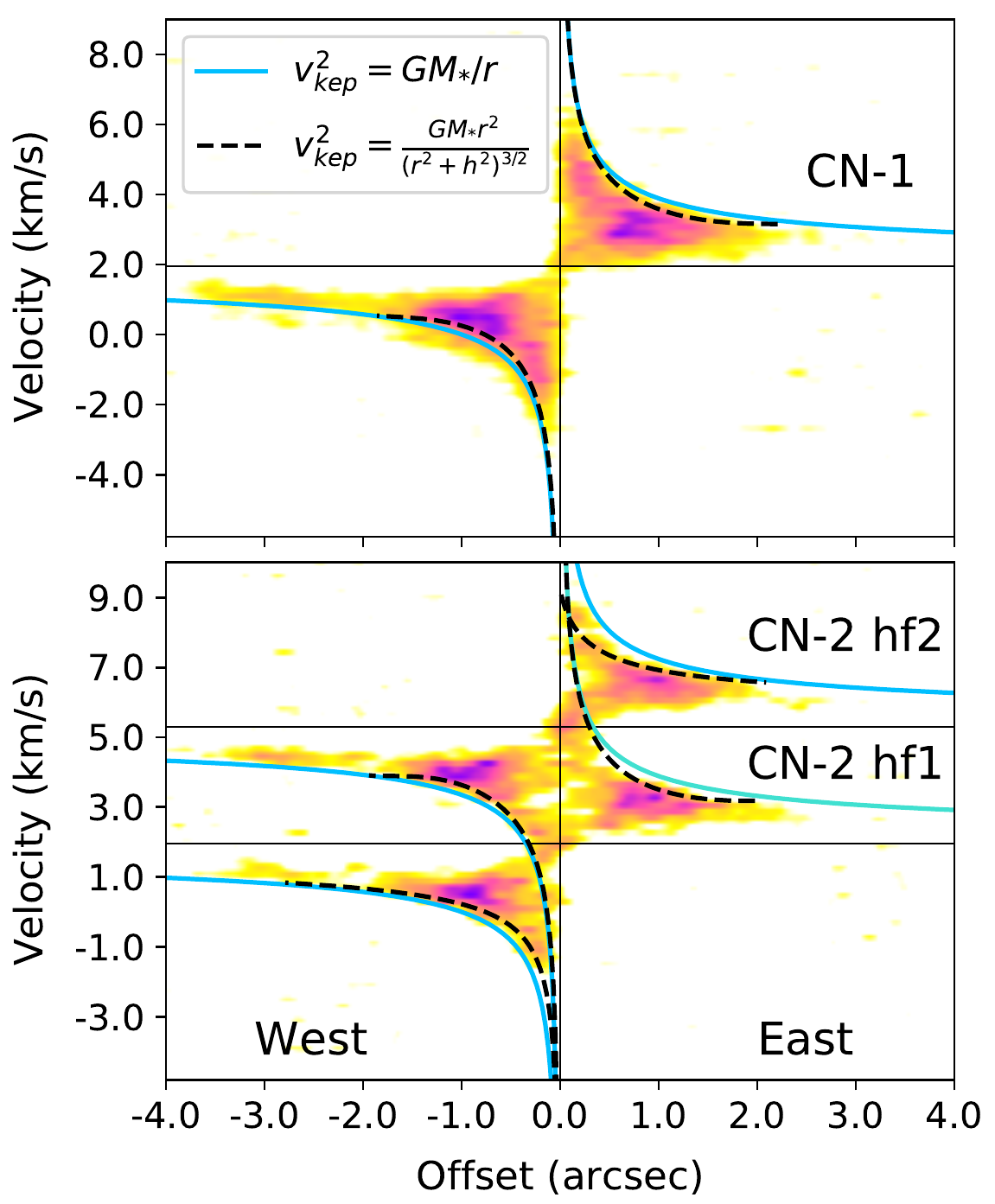}
      \caption{ Position-velocity diagrams along the semi-major axis of the disk (PA = 118.8$^{\circ}$) for CN-1 (top) and CN-2 (bottom, with two hyperfine group transitions). The offset is positive towards the east side of the disk and negative towards the west. The horizontal lines mark the systemic velocity ($\sim $ 1.95\,km\,s$^{-1}$), shifted for CN-2 hf2 emission to $\sim $ 5.3\,km\,s$^{-1}$. The blue curve marks the expected Keplerian velocity curve assuming a flat disk geometry. The dashed black line is the expected Keplerian velocity considering the derived best-fit model for the vertical profile of each tracer. Both models consider a stellar mass of 0.46$M_{\odot}$ as derived in \citet{veronesi_mass}. East and west sides of the disk, are marked in the bottom panel for reference.
              }
         \label{pvdiagram}
\end{figure}

Figure \ref{comp_CN_CO} presents the comparison of the derived vertical structures of each molecule, for the west side of the disk. As expected from disk chemical models \citep{2007prpl.conf..751B, 2014prpl.conf..317D}, the molecule that traces closest to the midplane is C$^{18}$O, followed by $^{13}$CO for most of the radial extent. The emitting surfaces of $^{12}$CO and CN transitions seem to be tracing almost the same layer up to $\sim$120\,au. The right panel of Figure \ref{comp_CN_CO} shows the $z/r$ measurements of each tracer. The mean values in each case are 0.53, 0.33 and 0.26 for $^{12}$CO, $^{13}$CO and C$^{18}$O, respectively. For the CN tracers the mean values across the radial extent for both tracers is 0.45. In the discussion (section 6) we analyse and compare this value to what has been derived for other Class II sources. Overall, our results are consistent with CN emission tracing the uppermost layers of the disk, which in Elias 2-27 seem to be significantly vertically extended.

We note that the radial extent of the derived vertical profiles is always smaller than the radial extent of the emission of each tracer. This can be seen from the selected CN channels, shown in Figure \ref{channels_CN_max}, where the black dots tracing the upper layer do not reach the border of the channel emission. In particular for $^{13}$CO and C$^{18}$O, the radial extent of the emission, as traced in the integrated intensity map, is $\sim650$ and $\sim$500\,au, respectively \citep{paneque_elias1}. Therefore the CO isotopologue's emission is roughly two times more extended radially than the furthest radial distance up to where we are able to trace the vertical structure. This occurs because we trace the upper layer of emission directly from the channel maps and only up to radial distances where we can separate upper and lower layers. At the border of the disk this is not possible. In the case of $^{12}$CO, which has the smallest radial extent, we are not able to trace further out due to the coarse spectral resolution and cloud absorption, which does not allow us to separate the emitting layers beyond $\sim$140\,au. As we are not able to sample further radial distances in the $^{12}$CO emission, the location of the ``turn-over'' in the height profile is not measured, contrary to the other molecules and it is unclear if at larger radii $^{12}$CO and CN would remain tracing similar locations.

\subsection{Lack of envelope or large scale emission}
\begin{figure*}[h!]
   \centering
   \includegraphics[scale=0.51]{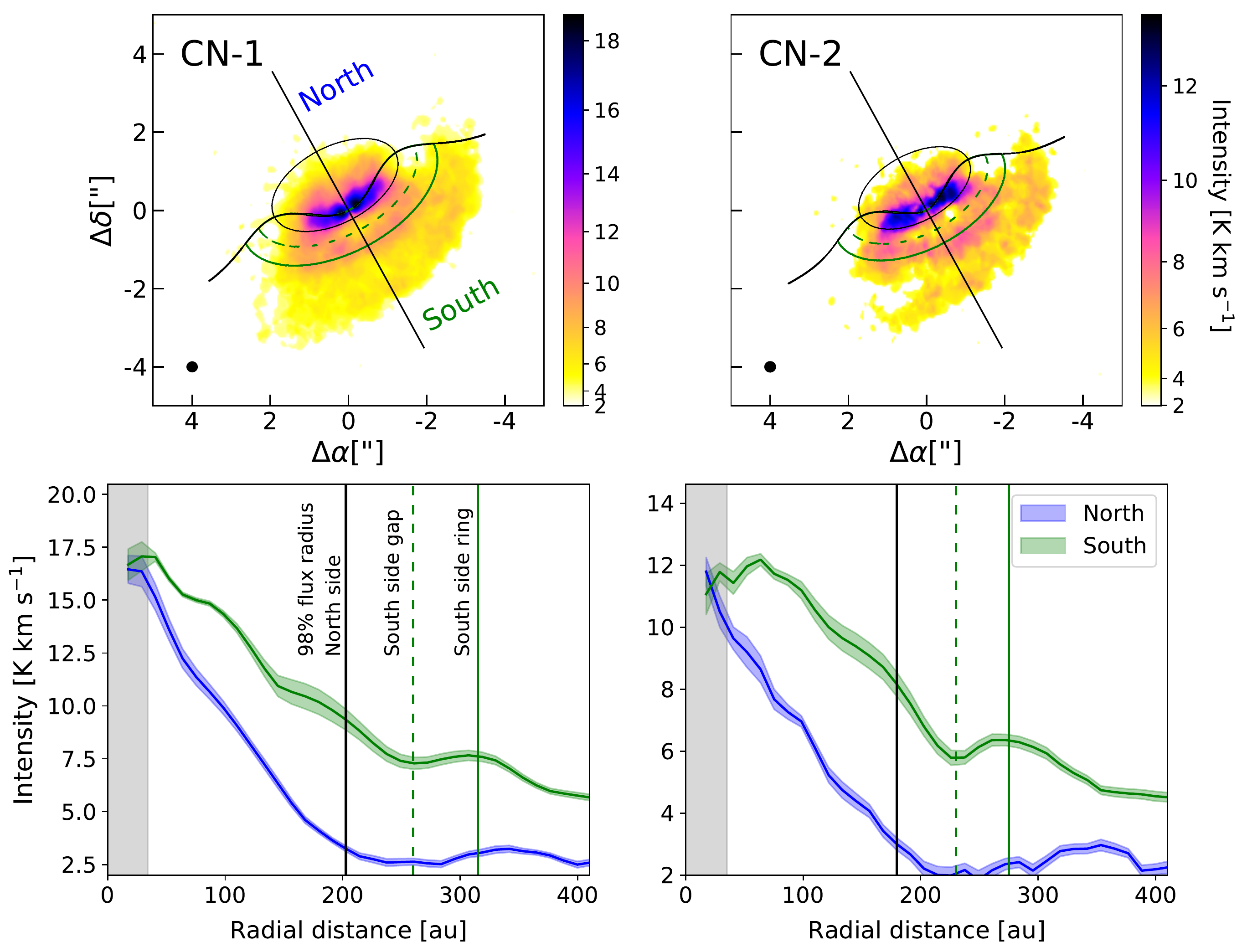}
      \caption{Top row: Integrated intensity of each CN transition, minor and major axis are marked considering the exponentially tapered vertical profile of the disk. The black elliptical contour marks the north quadrant's 98\% flux radius, derived from the intensity profile analysis. North and South regions are marked in the colors that are used for the intensity radial profiles. Dashed and continuous green contours indicate the location of the gap and ring (respectively) detected in the south side emission. Bottom row: Intensity profiles derived from the integrated intensity map of each transition. Blue curve traces the north side emission and green curve the south side emission. The vertical black lines show the location of the north emission 98\% flux radius. Vertical green lines show the location of the identified gap (dashed) and ring (solid) in the south side emission. The grey vertical shaded area marks a one-beam distance from the center. The colored shaded regions show the uncertainty of the intensity profile within each radial bin, obtained as the standard deviation of the data over the number of beams within the radial bin.
              }
         \label{mom0_intens_prof}
\end{figure*}
Considering the high $z/r$ values and vertical extension that we derive in Elias 2-27, we analyze the CN emission using a Position-Velocity (PV) diagram, to see if there is any indication of non-Keplerian velocities that may be associated to winds or other effects in the high disk atmosphere \citep{2006ApJ...646.1070A}. Figure \ref{pvdiagram} shows the PV diagram along the major axis of the disk, for each of the CN transitions. Overlayed is the expected velocity profile, considering also the best-fit model for the vertical profile and a stellar mass of 0.46$M_{\odot}$, as derived by \citet{veronesi_mass}. There is no indication of non-Keplerian motion and from the emission maps we have already discarded any large-scale emission (section 2), as seen in CO isotopologues \citep{paneque_elias1}, therefore the disk has intrinsically vertically extended CN emission. It is still plausible that winds or larger-scale structures are present in Elias 2-27, but, due to its high critical densities, CN may not be emissive enough to be detected.

\section{Asymmetric distribution of CN emission}

\begin{figure}[h!]
   \centering
   \includegraphics[scale=0.4]{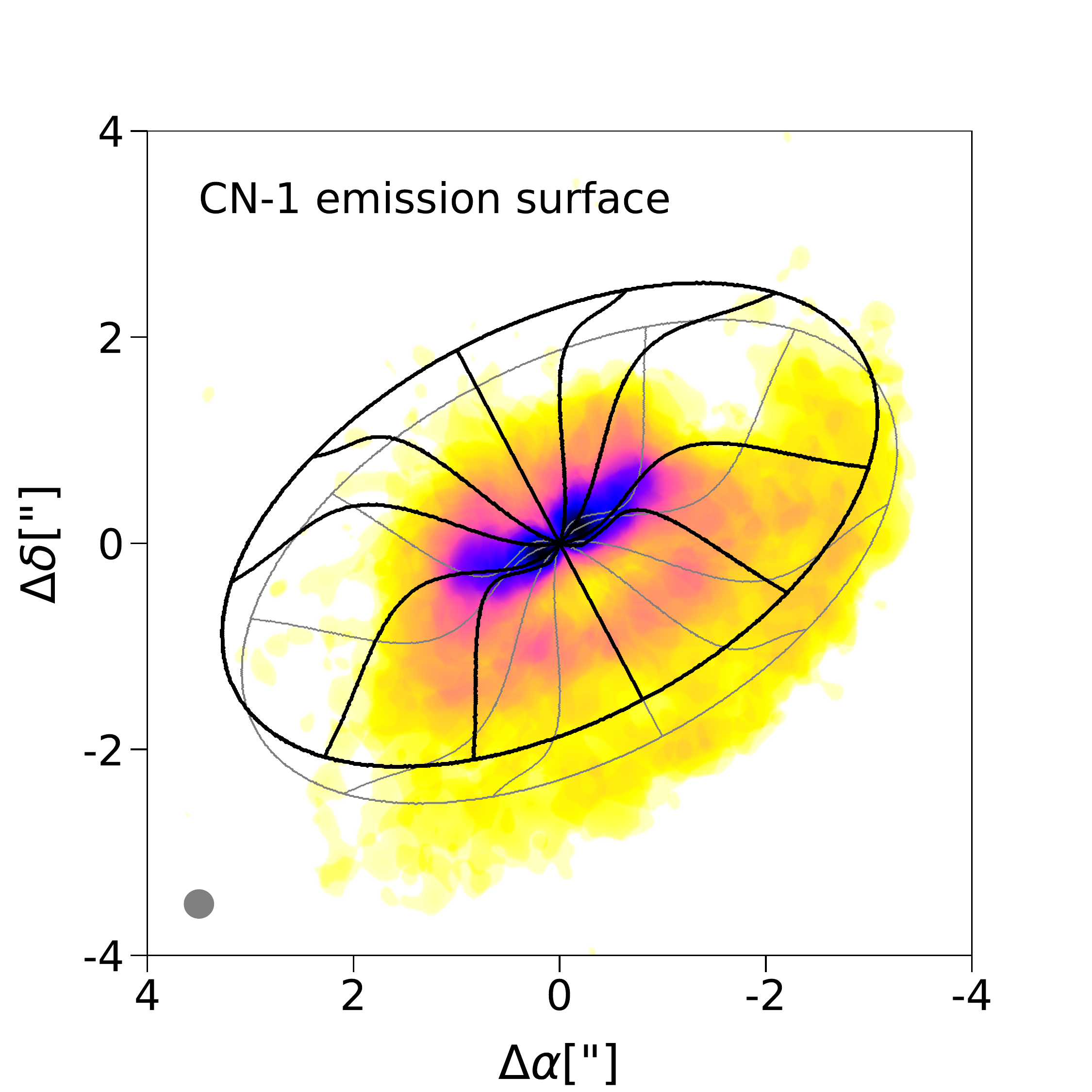}
      \caption{ Integrated intensity of CN-1. Overlayed are the emission surfaces using the exponentially tapered vertical profile and assuming symmetry between upper (black) and lower (grey) layers.
              }
         \label{model_surface}
\end{figure}

The most striking feature from the PV diagram analysis is the extent of the emission to the west side (negative offset), which is larger than in the east side (positive offset). This was commented in section 2 and relates to the extension difference seen between north and south regions of the disk, which is not perfectly symmetric in its azimuthal extent with respect to the disk major axis. In the south, the radially extended emission stretches azimuthally more towards the west than the east.

To study the north/south radial extent asymmetry, the emission of each tracer is adequately deprojected using the corresponding vertical structure previously derived (Figures \ref{height_CN_best} and \ref{height_CN_sides}). Azimuthally averaged intensity profiles and stacked spectra are obtained directly from the integrated intensity maps using the \texttt{get\_2D\_profile} function from \texttt{gofish} \citep{2019JOSS....4.1632T} and sampling in radial bins of 0.1$\arcsec$ (one third of our beam). The data analysis is separated between north and south sides of the disk, as defined by the major axis. Figure \ref{mom0_intens_prof} presents the integrated intensity maps and respective intensity profiles for each CN transition. We note that the analysis of CN-2 considers both hyperfine transitions.

The intensity profiles from the north side of the disk show lower values than the south side in both transitions, at all radii. For CN-1 the 98\% flux radius of the north side emission is 202.8$\pm$11.6\,au and for CN-2 179.6$\pm$11.6\,au (errors correspond to the radial bin size). Figure \ref{mom0_intens_prof} shows the respective radial contour of the 98\% flux radius overlayed on the integrated intensity map of each tracer. We do not resolve any structure from the north side intensity profiles, but beyond 300\,au there is a small bump of emission in both transitions. This bump is likely due to the ''tail'' of emission that extends slightly towards the north in the south-east side of the disk and can be appreciated in the integrated intensity maps (Fig. \ref{mom0_intens_prof}).

Theoretical \citep{2018A&A...609A..93C} and observational \citep{2020ApJ...899..157T, 2019A&A...623A.150V, MAPS_CN_Bergner} studies have shown that the radial profile of CN is expected to present a ring like feature which may become shallower or more peaked depending on the disk's physical properties \citep{2018A&A...609A..93C}. Indeed, the integrated intensity maps show a ``band-like'' emission in the south and we identify local maxima and minima from the south profile that we relate to the presence of, at least, one gap and one ring in each transition. The location of these structures are determined through scrutiny of the radial intensity profile and, when overlayed in the integrated emission maps, there is excellent agreement between the identified locations and the observed band-like structures (top row Fig. \ref{mom0_intens_prof}). For CN-1 the gap is located at 260\,au and the ring at 315\,au. In CN-2 we determine a position of 230\,au for the gap and 275\,au for the ring. The small differences in the location of gaps and rings between both transitions is likely due to the differences in the vertical profiles that are used for deprojecting the emission. We note that in the inner 200\,au there may also be tentative gaps and/or rings, however we are not able to visually assess them in the integrated emission maps with high confidence and do not classify them. The south profile also shows a possible dip of emission in the inner disk, which is expected for CN due to the high temperatures near the star that causes CN to be destroyed through its reaction with H$_2$ \citep{2018A&A...615A..75V}.

Figure \ref{model_surface} displays the expected emission surfaces for upper and lower layers from the best-fit exponentially tapered height profile, assuming they are symmetric. The maximum radial extent in both surfaces is 400\,au which is comparable to the sampled radii in the intensity profiles (Fig. \ref{mom0_intens_prof}). Figure \ref{model_surface} shows that the extended emission in the southernmost part of the disk is not explained by a projection effect of the lower layer, as it is expected to extend only slightly beyond the upper layer. A lack of emission in the north is also observed.

\section{Physical properties}

To measure the system properties and abundance parameters, one can compute rotational diagrams \citep{1999ApJ...517..209G}. We use this approach for the CO isotopologues and CN hyperfine group transitions.  The analysis of a rotational diagram relates the transition's theoretical properties, particularly the upper state energy, to the measured quantities obtained from the flux of the emission. If several transitions of the same tracer with varied upper state energies are available, this method allows an estimate of both the excitation (or rotation) temperature and the column density of the tracer. As only one transition of each molecule in CO and CN-1, together with the two CN-2 hyperfine group transitions which are too close in their upper energy values (see Table \ref{table_molec}) are available,  we are not able to simultaneously fit for the temperature and the total column density of each tracer as done in previous works \citep[e.g.,][]{1999ApJ...517..209G, 2018ApJ...859..131L, Facchini_2021}. Therefore, we fit only the total column density and use an estimate for the rotational temperature, as done by \citet{Facchini_2021}.

\begin{figure}[h!]
   \centering
   \includegraphics[width=\hsize]{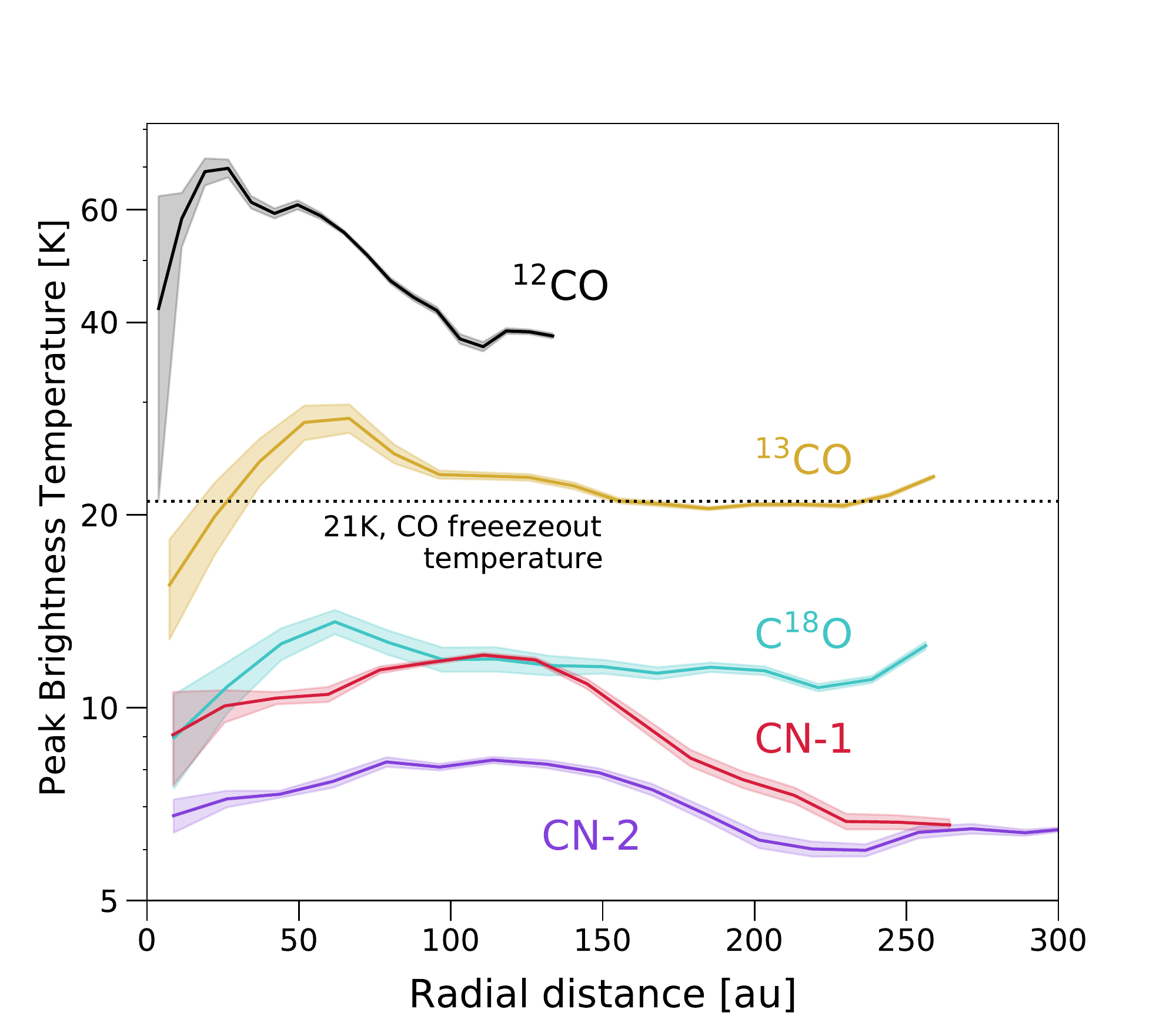}
      \caption{Azimuthally averaged peak brightness temperature profiles of each tracer available for Elias 2-27. The data are extracted only from the west side of the disk to avoid the high absorption effects of the east. In the case of CN we use data only from the south side, due to the additional north-south emission asymmetry. Shadowed colored regions mark the uncertainty in the azimuthal averaging of the peak brightness temperature maps, obtained by dividing the standard deviation by the number of independent beams in each radial bin area.
              }
         \label{temp_prof}
\end{figure}

For optically thick tracers, such as $^{12}$CO and $^{13}$CO, the brightness temperature will be coincident with the kinetic temperature. This approach has been used in previous works to determine the 2D radial and vertical temperature structure of disks \citep[][]{pinte_2018_method, MAPS_4_height_law, 2022ApJ...928....2I}. We have constrained that CN comes from a vertical layer, close to $^{12}$CO and higher than $^{13}$CO (Fig \ref{comp_CN_CO}), so we use the peak brightness temperature profiles of these two CO isotopologues as upper and lower limits for the kinetic temperature in the region from where CN emits. C$^{18}$O traces a layer just below $^{13}$CO, so the $^{13}$CO brightness temperature profile can be used as an estimate of the kinetic temperature of the region where C$^{18}$O emits. We use the on-line tool RADEX \citep{2007A&A...468..627V} to estimate the minimum H$_2$ number density needed for the rotation temperature to equal the derived kinetic temperature. For $^{12}$CO, $^{13}$CO and C$^{18}$O densities of order 10$^5$\,cm$^{-3}$ are sufficient. In the case of CN, the number density must be at least 10$^8$\,cm$^{-3}$. Models predict that at the location of CN emission the number density is 10$^6$-10$^8$\,cm$^{-3}$ \citep{2018A&A...609A..93C}. RADEX simulations show that these lower H$_2$ densities result in lower CN rotation temperatures of 20-40\,K assuming a kinetic temperature of $\sim$40\,K. These values are within the range that is sampled assuming the $^{12}$CO and $^{13}$CO temperature profiles as upper and lower limits (see Fig. \ref{temp_prof}).

The brightness temperature profiles are computed from peak brightness temperature maps and shown in Figure \ref{temp_prof}. Due to the asymmetries and cloud absorption, only the west side emission is considered for CO isotopologues and only the south side emission is considered for CN transitions. $^{12}$CO traces between 35-60\,K, $^{13}$CO has a peak temperature of $\sim$28\,K and C$^{18}$O varies with a mean temperature value of 11.9\,K. The CN tracers have mean values of 9.4\,K for CN-1 and $\sim$7.1\,K for CN-2. All mean brightness temperature values are calculated considering radial distances beyond 40\,au to avoid effects of beam smearing in the innermost radii. We emphasize that these profiles do not necessarily correspond to the kinetic and/or rotation temperature at the location of the molecules, this is only true for optically thick tracers. Indeed, $^{12}$CO and $^{13}$CO show high brightness temperatures, above the freeze-out temperature for CO ($\sim$21\,K, \citealt{2001A&A...377..566V, 2006A&A...449.1297B}). The low brightness temperatures of C$^{18}$O and both CN transitions are likely caused by the low optical depth of both molecules, which is why these values are not accurate representations of their kinetic temperature. Indeed, RADEX simulations of CN result in similar brightness temperatures (10.5\,K and 7.8\,K for CN-1 and CN-2, respectively) compared to our measurements when the column density of CN is $\sim$6$\times$10$^{13}$cm$^{-2}$ and the kinetic temperature 40\,K. We note that if the number density is very low in the upper disk layers, CN may be subthermally excited \citep{van_Zadelhoff_2003} which would also result in low brightness temperatures.

\begin{figure*}[h!]
   \centering
   \includegraphics[scale=0.48]{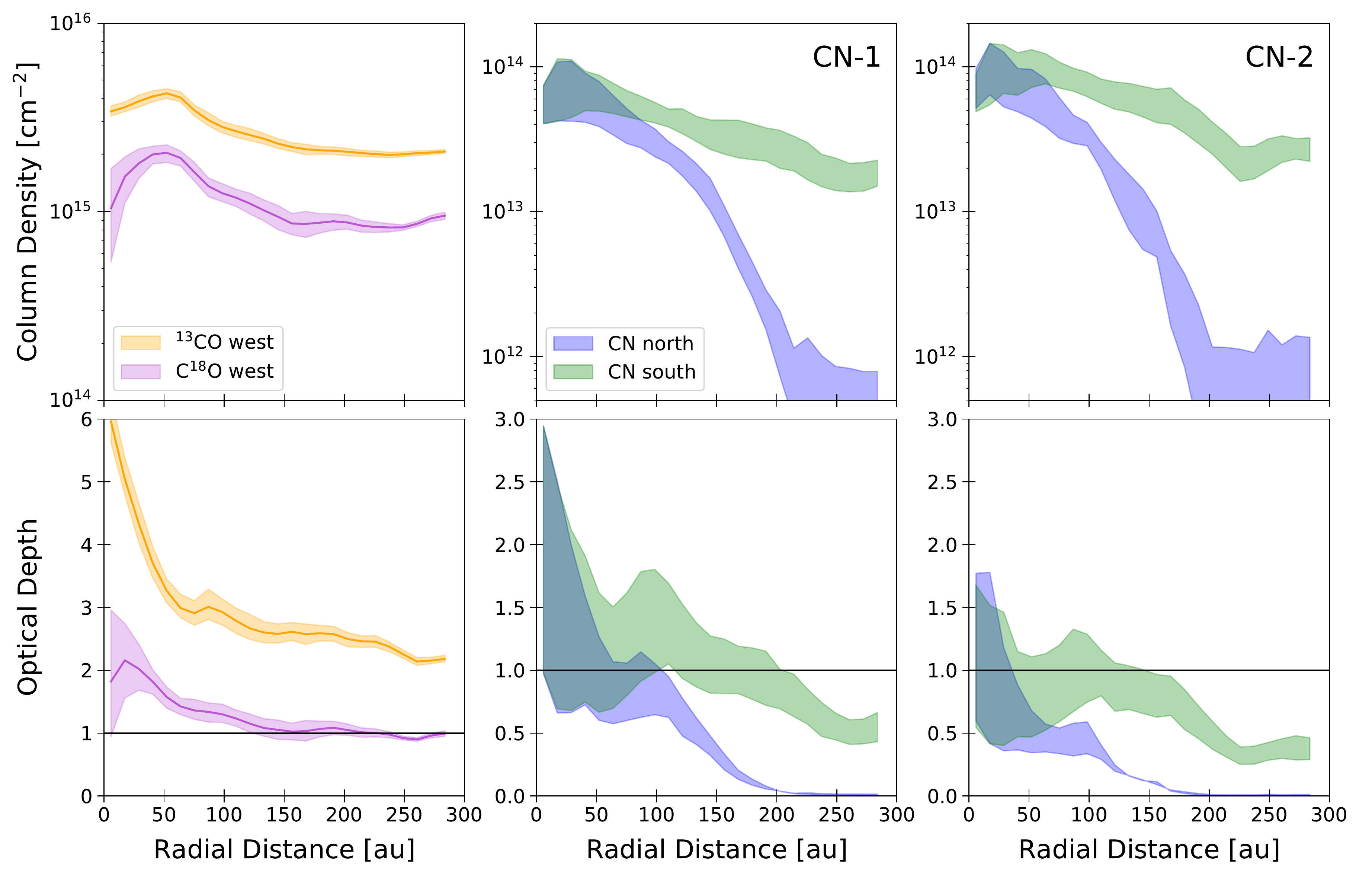}
      \caption{Top row: Column density profiles for each analysed tracer. Jointly in the leftmost panel $^{13}$CO and C$^{18}$O. Middle and Right panels show the column density profile of each CN hyperfine group, separated between north and south sides of the disk. Bottom row: optical depth profile for each molecule, following the same order as the top row. Horizontal black lines mark optical depth of $\tau = 1$. In all panels, the shaded regions show the uncertainty on the model.
              }
         \label{CO_CN_coldens}
\end{figure*}

With the temperature profiles of $^{12}$CO and $^{13}$CO as estimates of the kinetic temperature in the region where the molecules under study are located, we can obtain an estimate of the total column density and also calculate the optical depth profiles of each tracer. To do this, we must first calculate the column density from the upper state ($N_{u}$) of each transition using the integrated intensity profiles (obtained in Figure \ref{mom0_intens_prof}) and the parameters of each transition (see Table \ref{table_molec}). For optically thin emission ($\tau \ll 1$):

\begin{equation}
    N_u = \frac{4 \pi S_{\nu} \Delta \varv}{A_{ul}\Omega h c},
\end{equation}

where $S_{\nu} \Delta \varv$ corresponds to the integrated flux density and $\Omega$ is the area over which the flux was integrated. $S_{\nu} \Delta \varv$ is obtained from the integrated intensity profiles of Figure \ref{mom0_intens_prof}, where the profile has been derived from radial bins. The intensity value from the radial profile must be multiplied by the area of the radial annuli used in each bin and divided by the beam area to obtain the correct flux density units. Each radial annulus has an area $\Omega$. $A_{ul}$ is the Einstein emission coefficient of the transition, $c$ the speed of light and $h$ the Planck constant. If the optical depth ($\tau$) of the emission is non negligible then a correction factor $C_{\tau}$ must be applied to the upper state column density,

\begin{equation}
   C_{\tau} = \frac{\tau}{1 - e^{\tau}}.
\end{equation}

The column density of the upper state can then be related to the total column density ($N_{T}$), considering the correction factor, through:

\begin{equation}
   \frac{N_{u}}{g_{u}} C_{\tau} = \frac{N_{T}}{Q(T_{rot})} e^{-E_{u}/ k T_{rot}},
\end{equation}

where $g_u$ is the upper state degeneracy, $E_u$ is the upper state energy and $k$ is the Boltzmann constant. $Q(T_{rot})$ is the partition function and we estimate its value at each temperature by interpolating with a cubic spline from the values of the CDMS catalogue \citep{2001A&A...370L..49M} as has been done in \citet{Facchini_2021}. Finally, the optical depth of the emission may also be estimated from the upper state column density measurement following:

\begin{equation}
   \tau_{ul} = \frac{A_{ul} c^3}{8 \pi \nu^3 \Delta \varv} N_{u} (e^{h\nu / k T_{rot}} - 1).
\end{equation}

Here $\Delta \varv$ is the line width, which is calculated assuming the broadening from turbulence is negligible and calculating the thermal broadening from the temperature and the molecular weight of the molecule under study. $\nu$ is the line central frequency. All other parameters have been defined previously. Equation 5 allows us to express $C_{\tau}$ in terms of $N_u$ and therefore solve equation 4 for $N_T$, considering the optical depth properties of the molecules.

Figure \ref{CO_CN_coldens} shows the results for the total column density $N_T$ (top row) and optical depth (bottom row) profiles in $^{13}$CO, C$^{18}$O and CN tracers. The temperature profile used for the CO isotopologues is only the $^{13}$CO profile and the calculation considers the temperature profile uncertainties for the fit. For CN we considered a temperature range, where the upper limit is given by the  $^{12}$CO temperature profile and the lower limit by the $^{13}$CO temperature profile. As the $^{12}$CO temperature is only constrained up to $\sim$130\,au, we consider the last temperature value, 38\,K, as a constant value for the remainder of the radial extent.The assumption of constant temperature in the outer radii is based from the temperature profiles of $^{13}$CO and C$^{18}$O that display roughly constant temperature profiles from $\sim$100\,au outwards (see Fig. \ref{temp_prof}). For the calculation of the CN physical parameters, we do not consider the CO temperature profile errors, as they are within the sampled range of temperatures. 

Both CN tracers have column densities of 10$^{14}$\,cm$^{-2}$ in the inner disk that then drop to 10$^{13}$\,cm$^{-2}$ in the south side outer disk and even lower in the north. In the north side the column density has a steep decline, reaching values bellow 10$^{13}$\,cm$^{-2}$ beyond 150\,au. Likewise, the optical depth of the north side emission rapidly goes to values below unity in the inner 100\,au, indicating that the emission is optically thin. The south side of the disk displays a slowly radially decreasing column density which is always within the same order of magnitude, 10$^{13}$\,cm$^{-2}$. We note that CN-2 displays a localized decrease in the column density of the south side at $\sim$230\,au, which coincides with the location of the tentative gap reported in section 4. Regarding the optical depth values of the south side, they are close to unity and show a localized increase at $\sim$100\,au. Therefore, beyond $\sim$150\,au CN is optically thin in the whole disk, however, the emission may be marginally optically thick in the south side, particularly inwards of $\sim$100\,au.

For the CO isotopologue emission, our results trace a total column density of $\sim$10$^{15}$\,cm$^{-2}$ for $^{13}$CO and C$^{18}$O, however $^{13}$CO has a factor of 2-3 higher values. Studies have shown that CO column densities vary between systems depending on the stellar parameters, disk size and mass \citep{2016A&A...594A..85M}. Within an order of magnitude, our derived values are in agreement with previous observational and theoretical results \citep{2016A&A...594A..85M, 2018A&A...619A.113M, Facchini_2021}. The optical depth values of $^{13}$CO are above unity, corroborating our initial assumption that it is optically thick. C$^{18}$O shows marginally optically thin emission from $\sim$100\,au outwards.

\section{Discussion} \label{sec:discuss}

\subsection{Origin of CN emission and comparison to theoretical models}

The main formation mechanism of CN has been proposed to be the reaction of N with vibrationally excited H$_2$ pumped by UV radiation, making CN a tracer of the upper layers of a disk.  \citep{2018A&A...609A..93C, 2018A&A...615A..75V}. Our analysis of CN transitions in Elias 2-27 spatially locates the molecule in a constrained uppermost layer of the disk, with $z/r\sim$0.45 in all tracers. The disk is sufficiently flared that we can identify upper and lower layers of emission in the channel maps (see Fig. \ref{channels_CN_max}) and from there, directly trace the vertical surface of emission. To our knowledge, this is the first time that a direct measurement is done for CN emission in a disk of moderate inclination, where it is also possible to analyze the radial and azimuthal distribution of material. Previous works have relied on temperature estimates \citep[e.g., ][]{2020ApJ...899..157T} or have been conducted in edge on systems, lacking information on the azimuthal distribution \citep{2021A&A...646A..59R}.

A key feature predicted by models and detected in previous observations of CN is a shallow ring-like structure in the intensity and column density profiles of the disk that is not related to any gas or dust surface density features, but due to chemistry \citep{2018A&A...609A..93C, 2019A&A...623A.150V, 2020ApJ...899..157T, 2021A&A...646A..59R, MAPS_CN_Bergner}. Both of the analysed CN transitions show a ring of emission in the south side intensity profiles at 315\,au and 275\,au. Additionally, the optical depth profiles show a feature at $\sim$100\,au, where tentative bumps in the radial intensity profile may also be identified (see Fig. \ref{mom0_intens_prof}). All of these features are detected in the south side of the disk. From the north emission we do not retrieve any feature, even in the inner ($<$200\,au) region where there is strong CN emission.

CN emission is predicted to be optically thin, with typical column densities of $\sim$10$^{14}$\,cm$^{-2}$ and to be spatially constrained to a thin layer in the upper disk by physical-chemical models \citep{2018A&A...609A..93C, 2018A&A...615A..75V}. Our results confirm these predictions in most of the radial extension, however, in the inner $\sim$100\,au some emission may be marginally optically thick ($\tau = $ 1-2, see Fig. \ref{CO_CN_coldens}). This could be due to Elias 2-27 hosting a massive, gravitationally unstable, disk. In agreement with T Tauri models \citep{2018A&A...609A..93C}, the column density of CN in Elias 2-27 declines less than an order of magnitude in the south side of the disk for both transition groups. The rapid decline of column density in the north side is discussed in section 6.3 and we do not associate it with a chemical process, rather to an effect of the disk environment. Overall, our results of morphology and chemical/physical conditions are in agreement with theoretical predictions for CN emission originating from UV radiation. 

Elias 2-27 is likely a gravitationally unstable disk \citep{paneque_elias1} and studies focused on the chemistry of GI disks have suggested that CN may be a good tracer of GI spirals \citep{2015MNRAS.453.1147E}. CN emission originating from the spirals should be observed at the midplane, for it would be produced by the desorption of CN from the dust grains along the spirals. CN emits from tens of au above the midplane, therefore, we are not able to compare the dust continuum structure with the CN emission, for they are not spatially co-located and there is no signature of any emission from the midplane. This does not rule out self gravity or that there is no desorption in the midplane. It is possible that, if GI-originated CN emission is present, it is invisible due to the bright emission arising from the higher disk layers or lack of sensitivity in the outer disk. Another option is that, as the spirals are traced between 50-200\,au \citep{laura_elias, DSHARP_Huang_Spirals, paneque_elias1} and CN is marginally optically thick in this region, it is not possible to observe emission down to the midplane.

\subsection{Comparison with resolved emission in other systems}

CN has been detected in several Class II systems, both in spatially unresolved \citep[see surveys by, ][]{2015A&A...578A..31R, 2013A&A...549A..92G, 2016A&A...592A.124G} and resolved emission \citep[e.g.][]{2019A&A...623A.150V, 2020ApJ...899..157T, 2021A&A...646A..59R, MAPS_CN_Bergner}. There have been various indirect and direct methods employed to measure the vertical location and physical conditions of CN. Overall, there seems to be agreement on the CN conditions in Class II systems. Studies on TW Hya \citep{2014ApJ...793...55K, 2020ApJ...899..157T} and the Flying Saucer \citep{2021A&A...646A..59R} both characterise CN column densities between 10$^{13}$ - 10$^{14}$\,cm$^{-2}$. In the MAPS \citep{MAPS_1_oberg} dataset the CN column densities peak at 10$^{15}$\,cm$^{-2}$ for most disks (IM Lup, GM Aur, AS 209, HD 163296, MWC 480), however, the azimuthally averaged column densities are $\sim$10$^{13}$\,cm$^{-2}$ for most of the radial extent \citep{MAPS_CN_Bergner}. These column density values are in agreement with what we derive in Elias 2-27 (see section 5). MAPS also finds marginally optically thick CN emission in the inner disk \citep{MAPS_CN_Bergner}, in agreement with our results.

Even though the column density and optical depth values are in agreement with previous works and theoretical predictions, the distribution of the CN emission in Elias 2-27 is peculiar. In particular, the vertical extent is much higher than the estimated values for all other Class II sources. Indirect estimates using temperature profiles \citep[TW Hya, ][]{2020ApJ...899..157T} or UV-radiation models \citep[MAPS collaboration, ][]{MAPS_CN_Bergner} trace the emitting surface of CN at $z/r \sim$0.2. The CN region in the Flying Saucer disk \citep{2021A&A...646A..59R}, which is the only one to have been modelled directly, shows vertically symmetric CN emission arising from an intermediate molecular layer, lower than $^{12}$CO. \cite{2021A&A...646A..59R} also measure very low kinetic temperatures in the CN region ($\leq$15\,K) which they attribute to the low mass of the star (0.57 M$_{\odot}$) and extended CN emission, proposing deep X-ray radiation as the origin of the CN. It is interesting to note that Elias 2-27 has a comparable stellar mass of 0.46M$_{\odot}$ \citep{veronesi_mass} to the Flying Saucer, however a very different vertical structure.

In the case of Elias 2-27 we have CN arising from a high layer of $z/r \sim$0.45, indicating that this disk is more vertically extended than previously studied Class II systems.  The vertical distribution of CN in Elias 2-27 is similar to that of Class I source IRAS 04302+2247, where the CN vertical extent traces $z/r$ $\sim$ 0.4 \citep[see Fig. 2 and discussion within][]{2020A&A...642L...7P}. Unfortunately, CN observations in IRAS 04302+2247 are contaminated due to the continuum over-subtraction and absorption, therefore we do not have further information to compare with our results.

Comparison of the CO vertical structure in Elias 2-27 with other Class II disks shows that the mean $z/r$ values for Elias 2-27 are much larger than those found in the MAPS sample, for all CO isotopologues \citep{MAPS_4_height_law}. The differences in $^{13}$CO and C$^{18}$O values may be due to the higher transition ($J = 3-2$) used in this work, however, for $^{12}$CO the same transition as in MAPS is used ($J = 2-1$). A tentative relation between $z/r$ and CO radial extent has been proposed such that larger $z/r$ values relate to more extended disks \citep{MAPS_4_height_law, Law_2022}. Elias 2-27 has an extended CO disk, consistent with this trend, however this relation also finds that vertically extended CO disks ($z/r > 0.3$) have mean $^{12}$CO brightness temperatures $<$30\,K \citep{Law_2022}. As shown in section 5, the brightness temperature of $^{12}$CO varies between 35-60\,K. A larger sample of objects with varied disk and stellar properties is needed to determine if indeed Elias 2-27 is an outlier or if there are more systems with hot and vertically extended gas emission.

\subsection{Asymmetric CN, further evidence for infall or a warp?}

\begin{figure}[h!]
   \centering
   \includegraphics[width=\hsize]{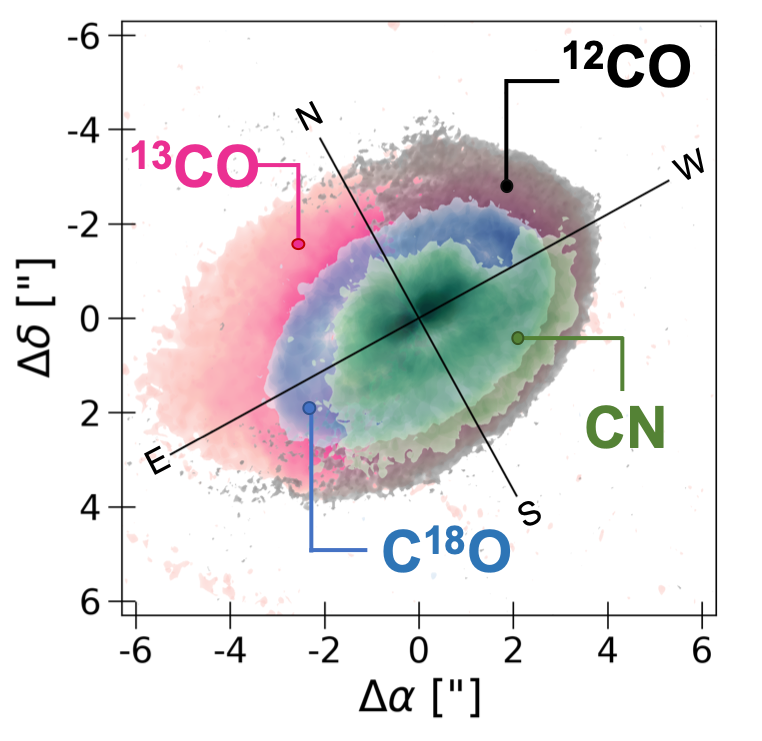}
      \caption{Integrated intensity maps for the tracers studied in this work, $^{13}$CO $J = $2-1, $^{13}$CO $J = $3-2, C$^{18}$O $J = $3-2 and CN $J =$ 7/2-5/2 (CN-1).
              }
         \label{elias_all_molec}
\end{figure}

The CN emission of Elias 2-27 shows an extreme asymmetry between the flux and density profiles in north and south sides of the disk. This was not seen in previous observations of CO emission \citep{paneque_elias1}, however it is similar to the morphology of available NACO L'-band observations \citep[see Fig. 12 in][]{2020A&A...635A.162L}. Figure \ref{elias_all_molec} displays the integrated intensity maps for the studied CO isotopologues and CN-1 emission for a comparison of the different radial and azimuthal locations. The east side of the disk shows radially extended CO emission and lack of $^{12}$CO, particularly in the north-east quadrant. We propose two scenarios to explain the CN asymmetric emission, that are also compatible with the various CO brightness and vertical extent asymmetries.

A decrease of CN production can be linked to presence of small grains in the north side disk atmosphere, which would shield the surface layers from FUV radiation and reduce the CN emission. An enhanced presence of small grains may be explained by high dust-to-gas ratio in this region \citep{2021ApJ...914..113N} or by infall of small grains. Infall is a mechanism proposed to be ongoing in Elias 2-27, due to the large-scale emission and variations between east/west sides observed in $^{13}$CO and C$^{18}$O \citep{paneque_elias1}. If material is still accreting onto the system, it would explain the vertical differences between east/west and also the north/south extension asymmetry seen in CN.

A second hypothesis to explain the asymmetry is the presence of a warp in the dust disk that creates a shadow and affects the FUV and X-ray radiation received by the north side of the disk \citep{2021MNRAS.505.4821Y}. A warped structure has been proposed to be responsible for the vertical asymmetries detected in the CO emission and the strong kinematical residuals \citep{paneque_elias1, 2022MNRAS.513..487Y, 2022ApJ...925..163R}. CN is expected to be a good tracer of the presence of a warp, through asymmetric distribution of its emission \citep{2021MNRAS.505.4821Y}. If there is a warped inner disk then we may characterize its inclination direction based on the south/north CN asymmetry and estimate it will be aligned with the disk major axis. The presence of a warp will not necessarily affect the CO abundance, however it is expected to affect the CO temperature. Indeed, previous studies have shown that the south side of the disk shows brighter CO emission \citep{laura_elias, paneque_elias1} which would be in agreement with shadowing of the north side by an inner warp \citep{2021MNRAS.505.4821Y}. With our current data we are unable to distinguish between the two (infall or warp) proposed scenarios.

Asymmetric CN emission, as reported in this work, has not been observed in any of the resolved Class II observations \citep{2019A&A...623A.150V, 2020ApJ...899..157T, 2021A&A...646A..59R, MAPS_CN_Bergner}, however, there is a similar case in the Class I protobinary source, Oph-IRS 67  \citep{2019A&A...627A..37A}. Oph-IRS 67 has CN emission traced only towards the south of the disk. This emission likely originates from a photon-dominated region (PDR) beyond the circumbinary disk \citep{2019A&A...627A..37A}. Besides CN, the asymmetry in Oph-IRS 67 is also traced in c-C$_3$H$_2$, DCN and H$_2$CO. It is possible that external radiation could also affect the emission of Elias 2-27. However we do not have any evidence of PDR regions beyond the disk that could affect the emission. Obtaining observations of PDR sensitive tracers could be useful to further study the asymmetry.

\section{Conclusions} \label{sec:conclude}

We present an extensive analysis of CN v$ = 0$, $N = 3 - 2$, $J = 7/2 - 5/2$ and $J = 5/2 - 3/2$ emission in Elias 2-27.  We also make use of previously published CO isotopologue observations to characterize the vertical stratification and physical properties (temperature, column density and optical depth) of the disk. Through this study Elias 2-27 becomes one of the few Class II systems with a detailed study of the vertical structure and chemical properties of its protoplanetary disk. Our main conclusions are the following:

\begin{enumerate}

    \item CN originates at a high vertical location from the disk midplane. It traces a region similar to $^{12}$CO, at least up to a radial distance of $\sim$150\,au. CN emits from a vertically constrained and mostly optically thin slab. This is in agreement with theoretical predictions of its main formation mechanism being the reaction of N with UV excited H$_2$

    \item We trace a strong CN emission asymmetry between the north and south sides of the disk. The extent of CN in the south side goes beyond 400\,au, while the 98\% flux radii of the north side is 180-200\,au. A ringed structure at 230-275\,au is observed in the south emission, as predicted by chemical models and similar to observations of CN in other systems. No structure is observed in the north side of the disk emission.
    
    \item The observed CN asymmetry may be explained by either ongoing infall or a warped inner disk. With the current data and models we are not able to confirm or reject either scenario.

    \item Elias 2-27 is a peculiar Class II source and the studied asymmetries share a strong resemblance with some Class I objects. It has vertically extended ($z/r >$0.3) CN and CO isotopologue emission, with high $^{12}$CO brightness temperatures and asymmetric CN distribution.
    
    \item Highly inclined sources such as Elias 2-27, are excellent systems to directly trace and analyse the disk structure and properties. Using optically thick tracers, such as $^{12}$CO and $^{13}$CO, we can trace the kinetic temperature structure. Combining the temperature profiles with information on the spatial location of various molecules we can recover the physical conditions of the gas.

\end{enumerate}

In summary, this study shows the vertical layering and structure of molecules in Elias 2-27. This is an excellent test-case for the importance of simultaneous analysis of multiple tracers to obtain constraints on temperature, column density and optical depth variations. The favorable inclination and high SNR make it possible to study the 3D disk structure so the emission can be traced radially, azimuthally and vertically.  Future observations with additional tracers, for example HCN that is UV and temperature sensitive, or c-C$_3$H$_2$ and DCN which can trace PDR regions, will allow us to fully study the vertical distribution and better understand the origin of the asymmetries in Elias 2-27.

\begin{acknowledgements}
This paper makes use of the following ALMA data: \#2013.1.00498.S, \#2016.1.00606.S and  \#2017.1.00069.S. 
ALMA is a partnership of ESO (representing its member states), NSF (USA), and NINS (Japan), together with NRC (Canada),  NSC and ASIAA (Taiwan), and KASI (Republic of Korea), in cooperation with the Republic of Chile. The Joint ALMA Observatory is operated by ESO, AUI/NRAO, and NAOJ. 
Astrochemistry in Leiden is supported by the Netherlands Research School for Astronomy (NOVA), and by funding from the European Research Council (ERC) under the European Union’s Horizon 2020 research and innovation programme (grant agreement No. 101019751 MOLDISK).
L.P. gratefully acknowledges support by the ANID BASAL projects ACE210002 and FB210003, and by ANID, -- Millennium Science Initiative Program -- NCN19\_171.
This project has received funding from the European Union’s Horizon 2020 research and innovation program under the Marie Sklodowska-Curie grant agreement No 823823, (RISE DUSTBUSTERS). 
\end{acknowledgements} 

\bibliographystyle{aa}
\bibliography{eliascn_paper.bib}

\begin{appendix}

\section{Channel Maps}

\begin{figure*}[h!]
   \centering
   \includegraphics[width=\hsize]{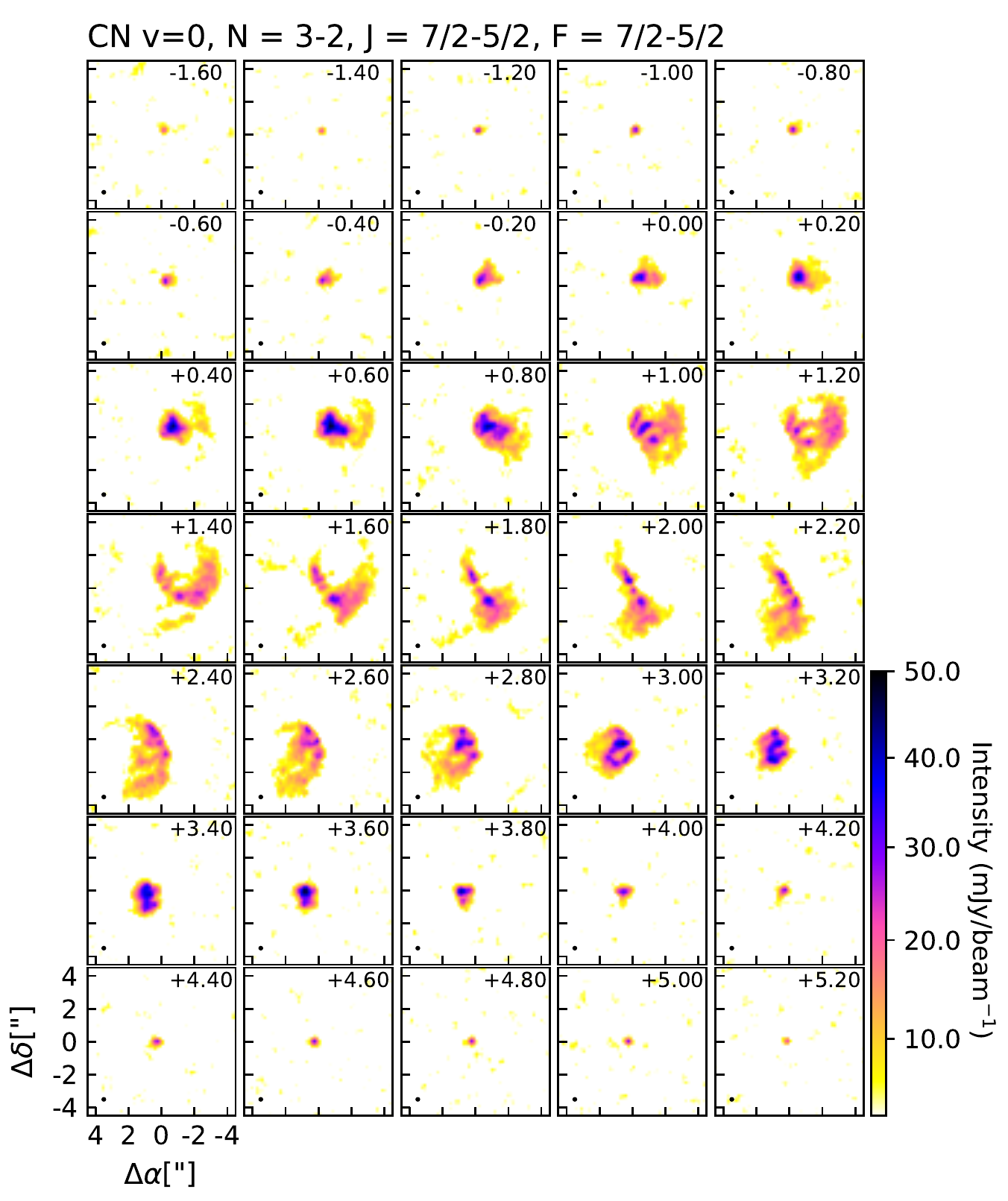}
      \caption{Complete channel map emission for CN $J = 7/2 - 5/2$. In each panel, the channel velocity is marked in the upper right corner and the beam size of the emission is shown by a black ellipse in the bottom left corner.
              }
         \label{panel_CN1}
\end{figure*}

\begin{figure*}[h!]
   \centering
   \includegraphics[width=\hsize]{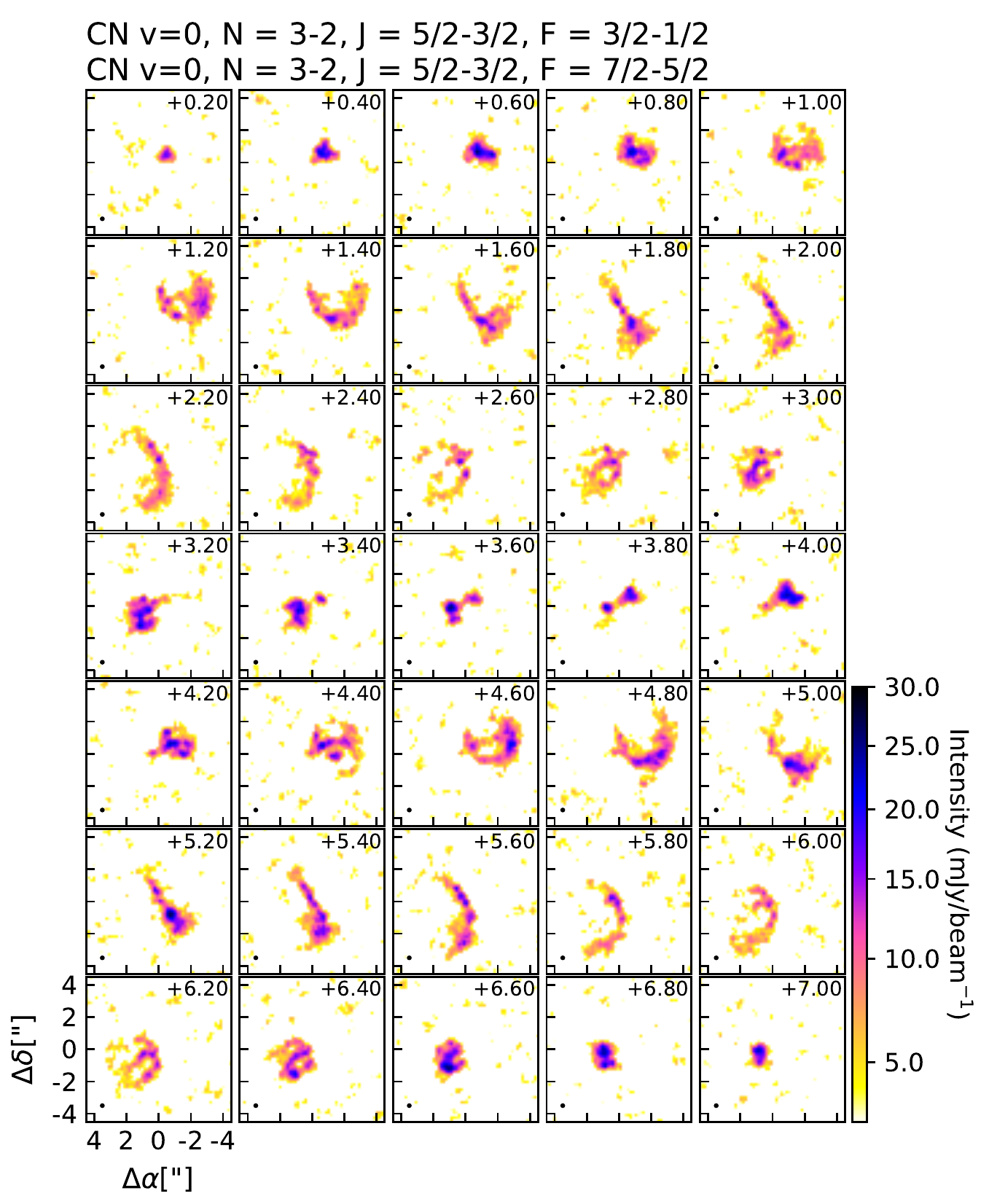}
      \caption{Complete channel map emission for CN $J = 5/2 - 3/2$. In each panel, the channel velocity is marked in the upper right corner and the beam size of the emission is shown by a black ellipse in the bottom left corner. This transition shows two consecutive hyperfine group transitions.
              }
         \label{panel_CN2}
\end{figure*}

In this section we present the complete channel maps for each of the CN transitions. Figure \ref{panel_CN1} shows the CN-1 emission, which has only one hyperfine group transition. Figure \ref{panel_CN2} shows the emission of the two hyperfine group structures of CN-2, imaged with respect to the frequency of CN-2 hf1, 340.0354080\,GHz \citep{2001A&A...370L..49M}. The systemic velocity of Elias 2-27 is estimated to be 1.95\,km\,s$^{-1}$ \citep{paneque_elias1}.

\section{CO isotopologue observations}

\begin{figure*}[h!]
   \centering
   \includegraphics[width=\hsize]{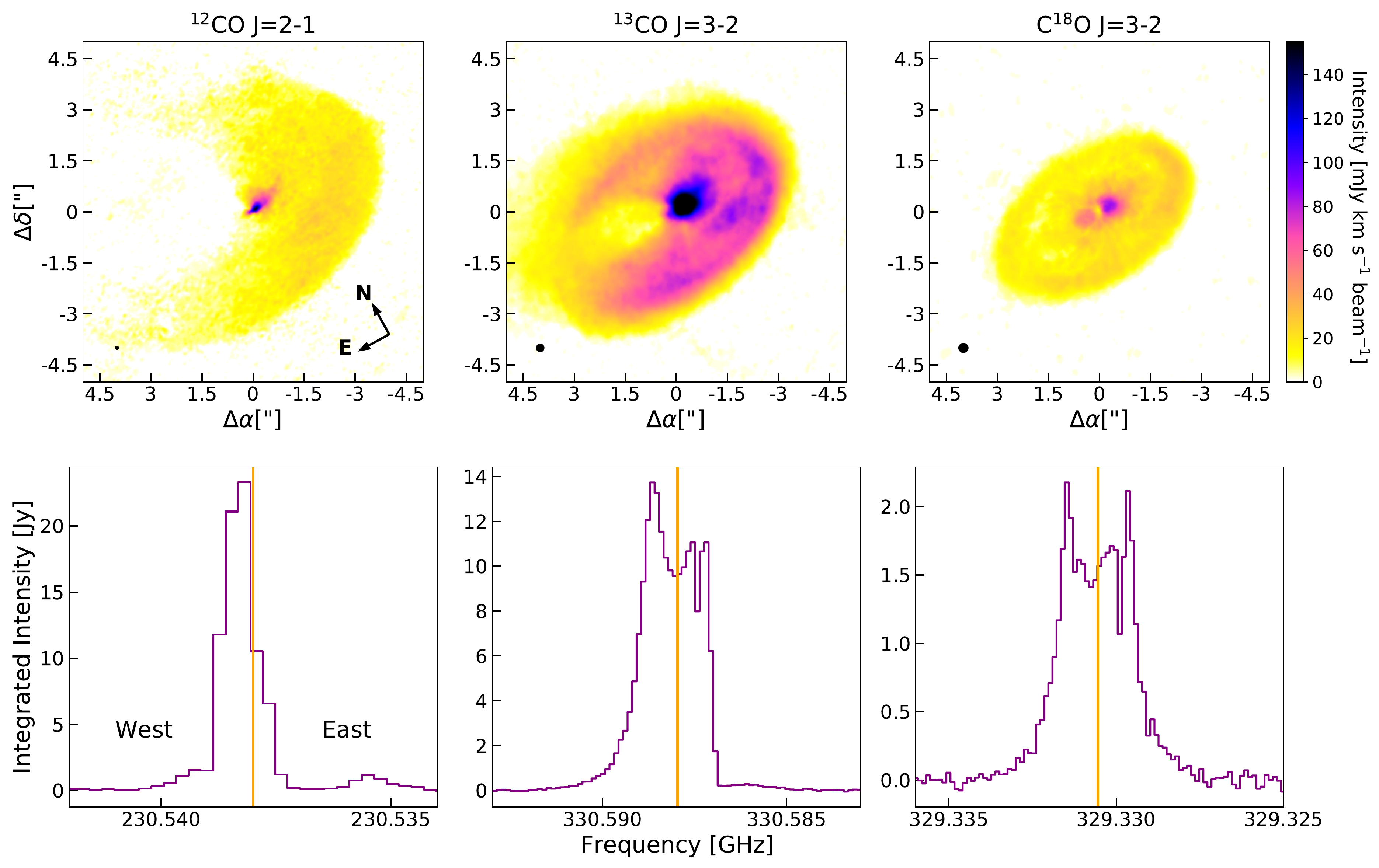}
      \caption{Top row shows the zeroth moment maps for each CO isotopologue used in this study. The beam size is shown in the bottom left corner of each panel. Bottom row displays the spectrum of each tracer. $^{12}$CO has a resolution of 0.35\,km\,s$^{-1}$ while $^{13}$CO and C$^{18}$O have a resolution of 0.111\,km\,s$^{-1}$. The orange vertical line marks the rest frequency in each case.
              }
         \label{obs_CO_spectra}
\end{figure*}

Figure \ref{obs_CO_spectra} shows the integrated intensity maps and spectra for the CO isotopologues used in this study. The detailed analysis of these data are presented in \cite{DSHARP_Andrews} and \cite{DSHARP_Huang_Spirals} for $^{12}$CO and in \cite{paneque_elias1} for $^{13}$CO and C$^{18}$O. The heavy contamination and absorption in the north east is visually identified in all tracers, least affecting C$^{18}$O.

We obtain the integrated fluxes and uncertainties of each tracer (shown in Table \ref{table_molec}) from the spectra in the emitting and line-free regions. For the total integrated flux we use an elliptical mask that encircles all of the emission, determined using a curve of growth method on the integrated intensity maps. This mask also determines the region from where the plotted spectra is derived (in Figures \ref{obs_CN_spectra} and \ref{obs_CO_spectra}). The uncertainties are derived from the standard deviation of the integrated flux retrieved from 20 line-free regions.

\end{appendix}
\end{document}